# Nanoscale electrothermal-switch superconducting diode for electrically programmable superconducting circuits

Tianyu Li[1,2,#], Jiong Li[3,#], Chong Li[1,4,#], Peiyuan Huang[1,5], Nuo-Zhou Yang[1,2], Wuyue Xu[1,2], Wen-Cheng Yue[1,5], Yang-Yang Lyu[1,*], Yihuang Xiong[1], Xuecou Tu[1,2], Tao Tao[5], Xiaoqing Jia[1,2], Qing-Hu Chen[3,*], Huabing Wang[1,2,*], Peiheng Wu[1,2], and Yong-Lei Wang[1,2,5,*]

[1] *Research Institute of Superconductor Electronics, Nanjing University, Nanjing 210023, China*

[2] *Purple Mountain Laboratories, Nanjing, 211111, China*

[3] *School of Physics, Zhejiang University, Hangzhou 310027, China*

[4] *School of Microelectronics, Nanjing University of Science and Technology, Nanjing, Jiangsu 210094, China*

[5] *State Key Laboratory of Spintronics Devices and Technologies, Nanjing University, Suzhou, China*

# These authors contributed equally

* Correspondence to: yylyu@nju.edu.cn; qhchen@zju.edu.cn; hbwang@nju.edu.cn; yongleiwang@nju.edu.cn

## Abstract

Superconducting diodes enable dissipationless directional transport, yet achieving electrical tunability and scalability remains a major challenge for circuit-level integration. Here, we demonstrate an electrothermal-switch superconducting diode in which a gate-controlled nanoscale hotspot dynamically breaks inversion symmetry in a superconducting nanowire. This mechanism gives rise to two coexisting nonreciprocal transport regimes—one associated with a nonreciprocal superconducting-to-normal transition and the other with ratchet-like vortex dynamics—both originating from the same electrothermal-switch process. The diode exhibits efficiencies up to 42% and 60% for the two regimes, respectively, and can be electrically switched on, off, or reversed in polarity in situ by applying a small gate current. These capabilities enable programmable superconducting circuits that realize electrically reconfigurable full-wave and half-wave rectification. The lithography-compatible design, high performance, and gate-controlled functionality establish a scalable platform for programmable superconducting electronics and hybrid quantum systems.

## Keywords

Superconducting diode; Superconducting nanowire cryotron; Nonreciprocal transport; Superconducting rectifier; Programmable superconducting circuits.

Superconducting diodes—devices that permit supercurrent flow preferentially in one direction without energy dissipation — have emerged as a powerful platform for exploring symmetry-broken superconducting states and for realizing low-power superconducting electronics and quantum technologies[1]. A variety of material platforms and device concepts have been proposed to realize the superconducting diode effect (SDE), including artificial superlattices[2,3], patterned superconducting films[4-8], ferromagnet-tuned devices[9-13], Van der Waals structures[14-19], twistronic stacks[20-24], and chiral superconductors[25,26]. In most cases, superconducting nonreciprocity arises from the combined breaking of spatial inversion symmetry and time-reversal symmetry. This is typically enforced through structural asymmetry, magnetic order, or spin-orbit coupling.

Despite these advances, achieving *in situ* and electrically controllable superconducting diode behavior remains challenging. Many existing implementations rely on external magnetic fields to tune the diode response, which limits local addressability, scalability, and circuit-level integration. Gate-tunable superconducting diodes have recently attracted growing interest[8,14,27-35], yet most designs are based on Josephson junctions that require delicate interface engineering and often face challenges in device uniformity and large-scale fabrication. Junction-free approaches, including van der Waals heterostructure[16], remain relatively rare and typically exhibit modest rectification efficiencies. Identifying alternative mechanisms that enable electrically tunable, scalable, and high-performance superconducting diodes therefore remains an important open problem.

Here we introduce an electrothermal-switch superconducting diode that addresses these challenges by exploiting gate-controlled local heating as a dynamic symmetry-breaking mechanism. The device is based on a cryotron-inspired superconducting nanowire architecture[36], in which a small gate current generates a localized nanoscale hotspot. Operating in the regime where the hotspot produces a controllable thermal gradient without fully quenching the nanowire, this electrothermal asymmetry breaks spatial inversion symmetry *in situ*, enabling electrically switchable nonreciprocal superconducting transport without modifying the magnetic field.

This electrothermal-switch mechanism gives rise to two distinct nonreciprocal transport regimes within a single device. At higher bias currents, the diode effect originates from a nonreciprocal superconducting-to-normal transition, while at lower dissipation levels it arises from asymmetric vortex dynamics associated with biased vortex entry and exit. Both regimes share a common electrothermal origin, providing a unified framework for understanding gate-controlled superconducting nonreciprocity. Beyond establishing this mechanism, the device architecture allows direct electrical control over diode polarity and supports reconfigurable superconducting circuits compatible with standard lithographic fabrication.

The SDE emerges when both spatial inversion symmetry and time-reversal symmetry are broken. In our device, inversion symmetry is controllably broken by a nanoscale hotspot generated through a gate current in a superconducting nanowire, producing a local thermal gradient (Figure 1a). The design is inspired by the superconducting nanowire cryotron (nTron) — a three-terminal superconducting device[36] that utilizes localized heating from a small gate current to modulate superconductivity. Owing to its ultra-compact footprint, immunity to magnetic noise, high fan-out capability, and compatibility with large-scale fabrication [37-39], the nTron has been deployed in diverse superconducting platforms [40-43], including single-flux quantum circuits[39], CMOS-hybrid memories[40], large-scale digital circuits[37], and single-photon detectors[42]. Despite these advances, its potential for generating nonreciprocal superconducting transport has remained unexplored.

Our devices were fabricated on Si substrates using patterned 10 nm thick NbN thin films, a material widely adopted in nTron-based circuits[39,40], single-photon detectors[42], superconducting microwave components[44], spiking neural networks[45], and full-wave bridge rectifiers[4]. In a conventional nTron, the nanowire width (<100 nm) ensures complete channel quenching when the gate current induces a hotspot. To create a controllable thermal gradient without fully driving the channel into the normal state, we increased the nanowire width to 500 nm (Figure 1b). Two narrow (~40 nm) gate leads were symmetrically placed on opposite sides of the nanowire to generate localized hotspots under small gate currents (Figures 1a and 1b). The four-terminal configuration allows the position of the hotspot—and therefore the direction of inversion-symmetry breaking—to be *in situ* switched by selecting which gate is biased. Detailed fabrication procedures are described in Supporting information.

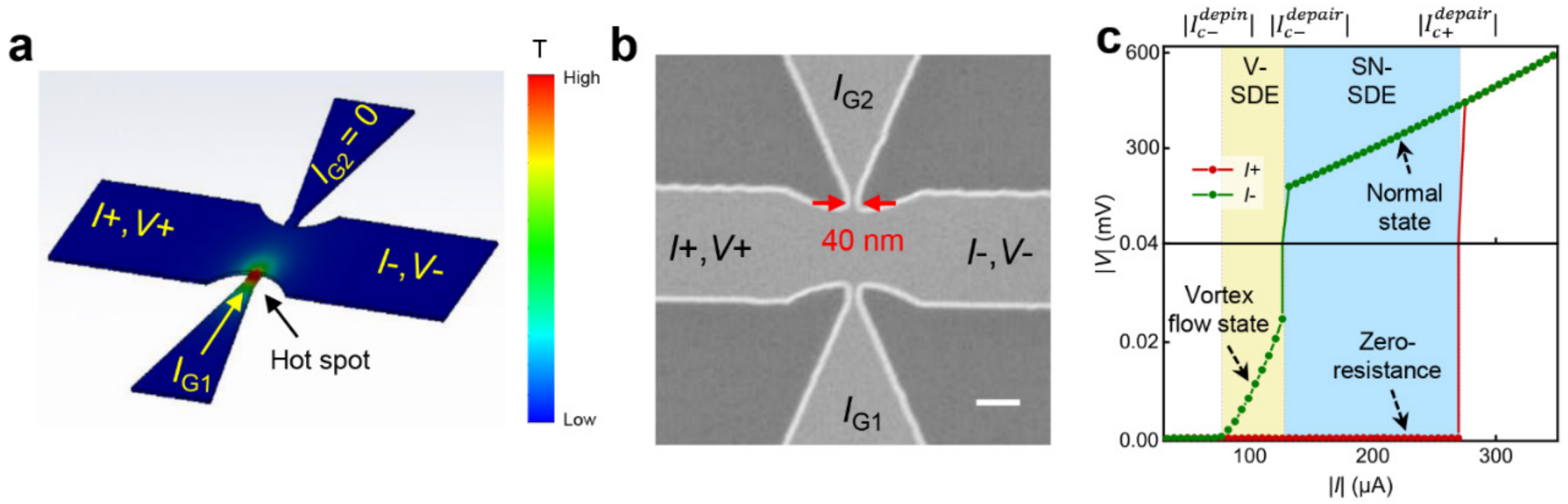

**Figure 1.** Electrothermal–switch superconducting diode. (a) Schematic of the electrothermal-switch superconducting diode, where a nanoscale hotspot is generated by a small gate current. The resulting thermal gradient locally breaks the spatial inversion symmetry of the superconducting nanowire channel. (b) Scanning electron microscope (SEM) image of a representative device. Two gate leads are symmetrically placed on the upper and lower sides of the main channel. Scale bar, 150 nm. (c) *I-V* characteristics measured with a gate current $I_{G1}$=10 μA applied to one gate lead at 2 K and an external magnetic field of 300 Gs. The top panel shows the high-dissipation regime corresponding to the nonreciprocal superconducting-to-normal transition, the bottom panel presents a magnified view, highlighting the low-dissipation regime from ratchet-like vortex motion.

Temperature-dependent resistance measurements revealed a critical temperature $T_c$=9.1 K (Figure S1a). With no gate current, the current-voltage (*I-V*) characteristics were symmetric, exhibiting negligible nonreciprocity even under applied magnetic fields (Figure S1b). When a small gate current $I_{G1}$=10 μA was applied to the lower gate lead, a localized hotspot formed on the lower side of the nanowire (as illustrated by Figure 1a), breaking inversion symmetry and inducing pronounced nonreciprocal transport, as evidenced from the markedly divergent *I-V* curves for opposite current directions (top panel of Figure 1c). Only the forward current-sweep curves, which determine the intrinsic critical currents, are shown here; the corresponding reverse-sweep data, used to extract symmetric retrapping currents, are presented in Figure S2.

The chosen gate current slightly exceeds the retrapping current of the gate lead ($I_r$=9.1 μA), marking the transition from the resistive to the superconducting state (Figure S3). Operating slightly above $I_r$ generates a stable nanoscale resistive hotspot without fully suppressing superconductivity in the main nanowire. In contrast, excessive gate currents cause overheating and completely quench superconductivity, eliminating the inversion-symmetry-broken state necessary for the SDE. Notably, the observed nonreciprocity is insensitive to the polarity of the gate current: reversing its direction leaves the diode characteristics unchanged (Figure S4). This confirms that the SDE originates from electrothermal switching, since Joule heating depends on the current magnitude rather than its direction. Furthermore, the absence of a clear diode effect for gate currents below the critical value (Figure S5) confirms that the hotspot, rather than the current bias itself, is the essential source of the observed nonreciprocity. These results establish an electrothermal-switch superconducting diode that can be electrically turned on or off *in situ* by a minute gate current.

To further understand the origin of this electrically induced nonreciprocity, we systematically analyzed the current–voltage characteristics in Figure 1c. Interestingly, two distinct nonreciprocal regimes emerge in the device (yellow and blue shaded regions in Figure 1c). The first regime (blue) clearly displays a diode effect arising from a nonreciprocal superconducting-to-normal transition, characterized by markedly different depairing currents for opposite bias directions ($I_{c-}^{depair}$= 137.5 μA, $I_{c+}^{depair}$= 269.5 μA). Upon magnification (bottom panel of Figure 1c), an additional low-dissipation regime (yellow) becomes evident, featuring voltages three orders of magnitude smaller than the normal-state resistance. This weaker dissipation arises from vortex dynamics—motion of quantized magnetic flux lines penetrating the type-II superconductor under an applied field. The onset of vortex motion defines the depinning current ($I_{c-}^{depin}$= 82.5 μA), below which vortices remain pinned and the nanowire retains a zero-resistance state. In the following, we refer to these two regimes as the superconducting-to-normal transition SDE (SN-SDE) and the vortex-motion SDE (V-SDE), respectively.

To further confirm that the observed dual-regime nonreciprocity originates from the electrothermal dynamics rather than structural asymmetry, we performed time-dependent Ginzburg-Landau (TDGL) simulations (see Methods)[5]. The simulated *I-V* characteristics successfully reproduce the experimentally observed two-stage nonreciprocity (Figure 2a). Video S1, with representative screenshots shown in Figures 2b and 2c, illustrates the spatial evolution of the superconducting order parameter (Cooper-pair density) under opposite current directions. When a magnetic field is applied, vortices penetrate the superconducting nanowire. Their motion dissipates energy and generates finite resistance, even while the sample remains globally superconducting. According to the Bean-Livingston model, the nanowire edges act as potential barriers for vortex entry and exit [46]. The gate-induced hotspot locally reduces the superconducting condensation energy, thereby lowering the edge barrier and creating an easy-entry pathway for vortices (Figures 2b and 2c).

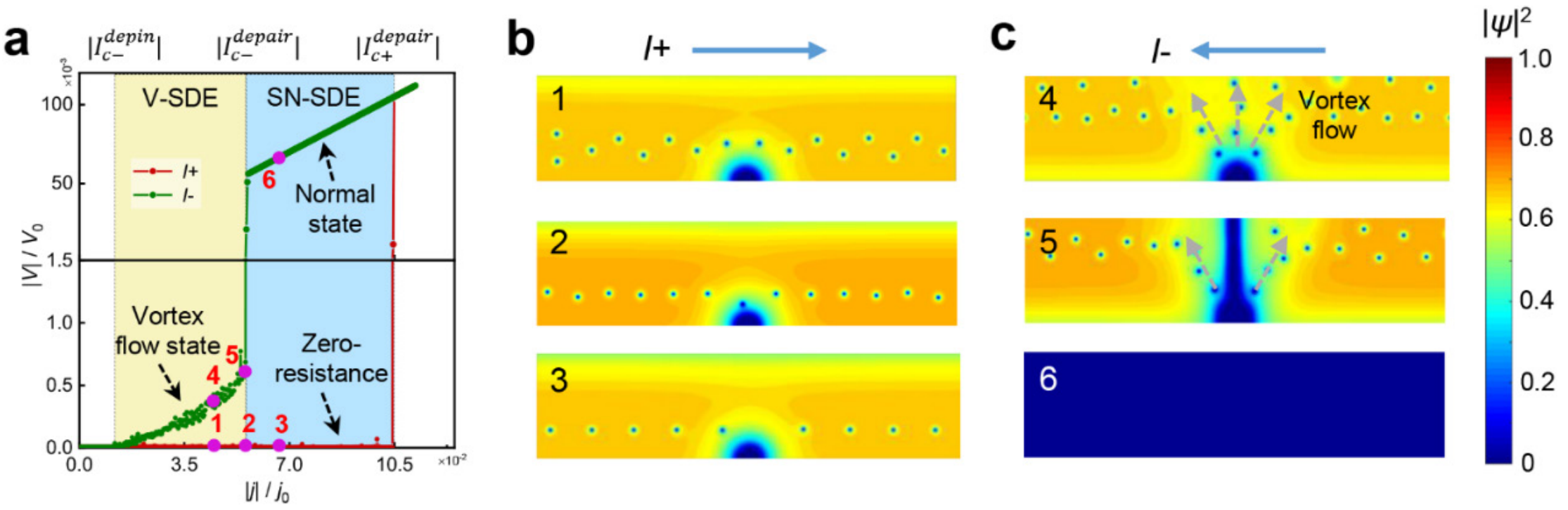

**Figure 2.** Microscopic mechanism of the superconducting diode effect revealed by TDGL simulations. (a) Simulated *I-V* characteristics showing two distinct nonreciprocal regimes, consistent with experimental observations. $j_0$ and $V_0$ are the dimensionless units of TDGL, for details, see the Methods. (b) and (c) Snapshots from TDGL simulations (Video S1) illustrating the spatial distribution of the superconducting order-parameter under positive (b) and negative (c) currents. The snapshots correspond to the *I-V* points (1-6) indicated in (a).

For positive current, the Lorentz force drives vortices from the upper edge toward the lower one, allowing them to exit freely through the hotspot, while the intact upper barrier prevents further vortex entry (Video S1; Figure 2b). The system therefore maintains a zero-resistance state up to a large depairing current amplitude, $|I_{C+}^{depair}|$. For negative current, by contrast, the Lorentz force injects vortices through the hotspot, establishing a vortex-flow channel that produces a finite resistance corresponding to the V-SDE regime. As the magnitude of the negative current exceeds $|I_{C-}^{depair}|$, enhanced Joule heating completely quenches superconductivity in the channel (Video S1; Figure 2c), giving rise to the high-voltage SN-SDE regime. Simulated temperature maps (Video S2; Figure S6) and electrothermal control results (Figure S7) corroborate these sequential transitions. These results reveal that both SN-SDE and V-SDE arise from a unified electrothermal-switch mechanism: gate-controlled hotspots concurrently modulate vortex dynamics and local superconductivity, leading to electrically tunable nonreciprocal transport.

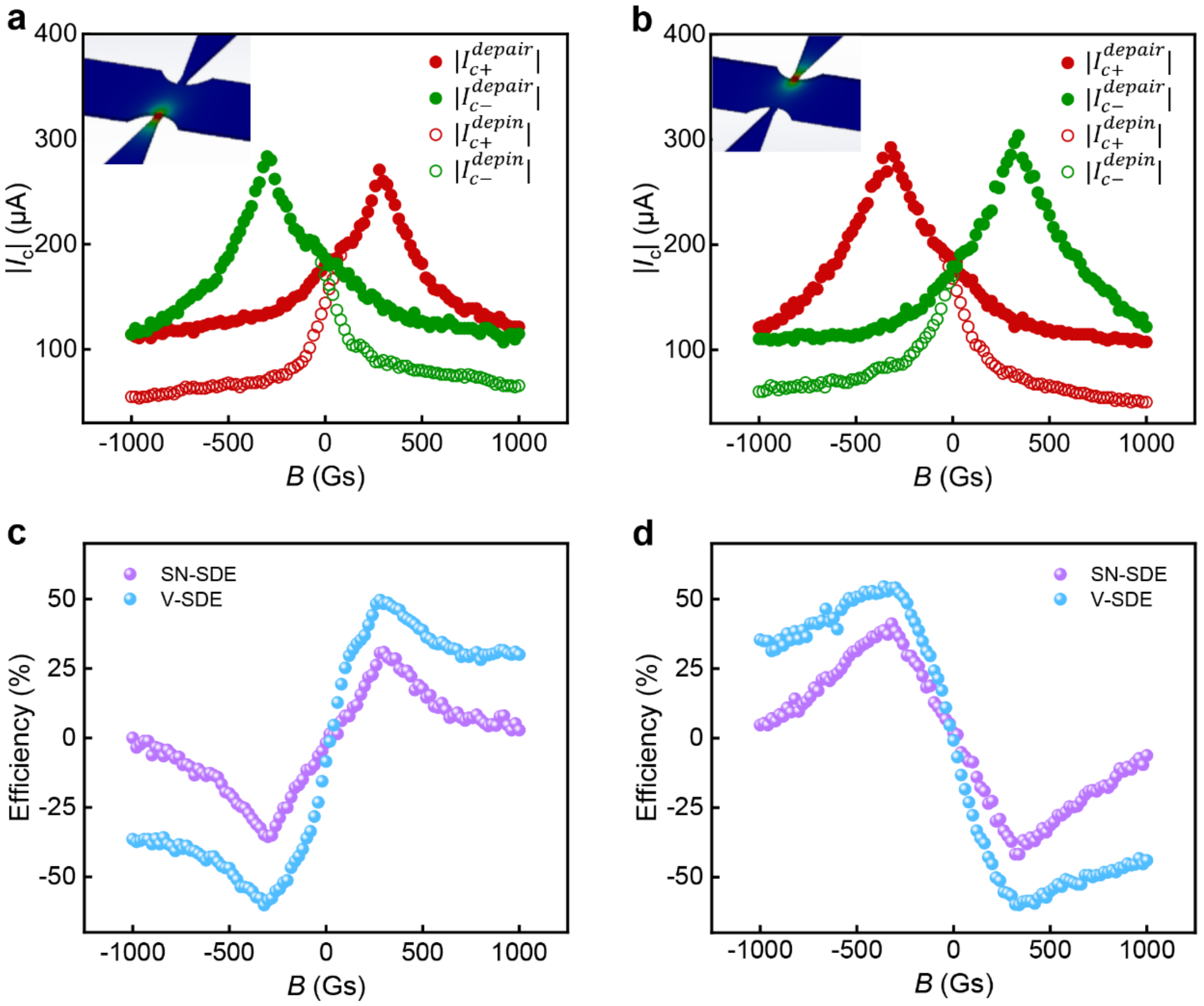

**Figure 3.** High efficiency and electrical tunability of the electrothermal-switch superconducting diode. (a) and (b) Magnetic-field dependence of the depairing and depinning critical currents measured for opposite current directions, with a gate current $I_{G1}$= 10 μA applied to the lower gate (a) and $I_{G2}$= 10 μA applied to the upper gate (b). Insets schematics illustrate the corresponding hotspot locations. (c) and (d)**,** Diode efficiencies corresponding to panels in (a) (c) and (b) (d)**.**

Achieving high diode efficiency in a scalable superconducting platform is crucial for circuit-level applications. Our lithographically defined electrothermal-switch diodes satisfy this requirement, combining robust performance with inherent scalability. Figure 3a shows the magnetic-field dependence of the depairing and depinning currents for opposite current directions, while Figure 3c presents the corresponding diode efficiencies, defined as $\eta = \frac{|I_{c+}|-|I_{c-}|}{|I_{c+}|+|I_{c-}|} \times 100\%$, where the sign of $\eta$ denotes the diode polarity. Under zero magnetic field, the diode effect is absent even with an active gated

hotspot (Figure S8). The diode efficiency exhibits a maximum at a magnetic field of approximately 320 Gs, beyond which it decreases. This behavior is determined by the interplay between the hotspot-induced spatial symmetry breaking and the Meissner screening currents, following the same mechanism discussed in Ref. 10. For the current direction with the larger critical current, only the depairing current $|I_c^{depair}|$ is observed, whereas the opposite direction exhibits both depinning and depairing currents, with $|I_c^{depair}| > |I_c^{depin}|$. Consequently, the efficiency of the V-SDE consistently exceeds that of the SN-SDE (Figure 3c). The maximum efficiencies reach 42% for SN-SDE and 60% for V-SDE—remarkably high values that reflect strong nonreciprocity and stable rectification (Figure S9).

The scalability and reproducibility of these diodes enable the realization of multi-element superconducting circuits. To demonstrate an electrically tunable circuit, we fabricated a full-wave bridge rectifier consisting of four identical electrothermal-switch diodes (Figure 4a)[4,6]. Configuring the diode polarities as illustrated in Figure 4b yields full-wave rectification (Figure 4h, an enlarged view of the data shown in Figure S10) under an applied alternating current drive (Figure 4g). The operational status of each diode during different temporal intervals is illustrated in Figure S11 and Table S1. Beyond magnetic-field modulation—where diode polarity can be reversed by changing the applied field (Figures 3a and 3c)—the electrothermal-switch diode offers direct, local, and reversible electrical tunability. As shown in Figures 3b, 3d and Figure S9, applying a gate current to the opposite-side electrode relocates the hotspot, thereby inverting the direction of inversion-symmetry breaking and switching the diode polarity. Consequently, the polarity of the full-wave bridge rectifier can be reprogrammed *in situ* without adjusting the magnetic field (Figures 4c and 4i).

This electrical programmability also enables independent control of individual diodes within a circuit, in contrast to global magnetic-field schemes where all elements respond collectively [4,6]. For the same bridge circuit (Figure 4a), setting the gate currents of two selected diodes to zero deactivates their nonreciprocity, converting the bridge into a half-wave rectifier (circuit schematic in Figure 4d; corresponding data in Figure 4j). The deactivated diodes behave as ordinary superconducting wires with higher, direction-independent critical currents (Figure S12). The half-wave rectifier polarity can also be reversed by reconfiguring the diode states (Figures 4e and 4k). Thus, the ability to electrically switch on/off and reverse diode polarities *in situ* enables multifunctional superconducting circuits—a level of programmability previously unattainable in architectures constrained by global magnetic-field control.

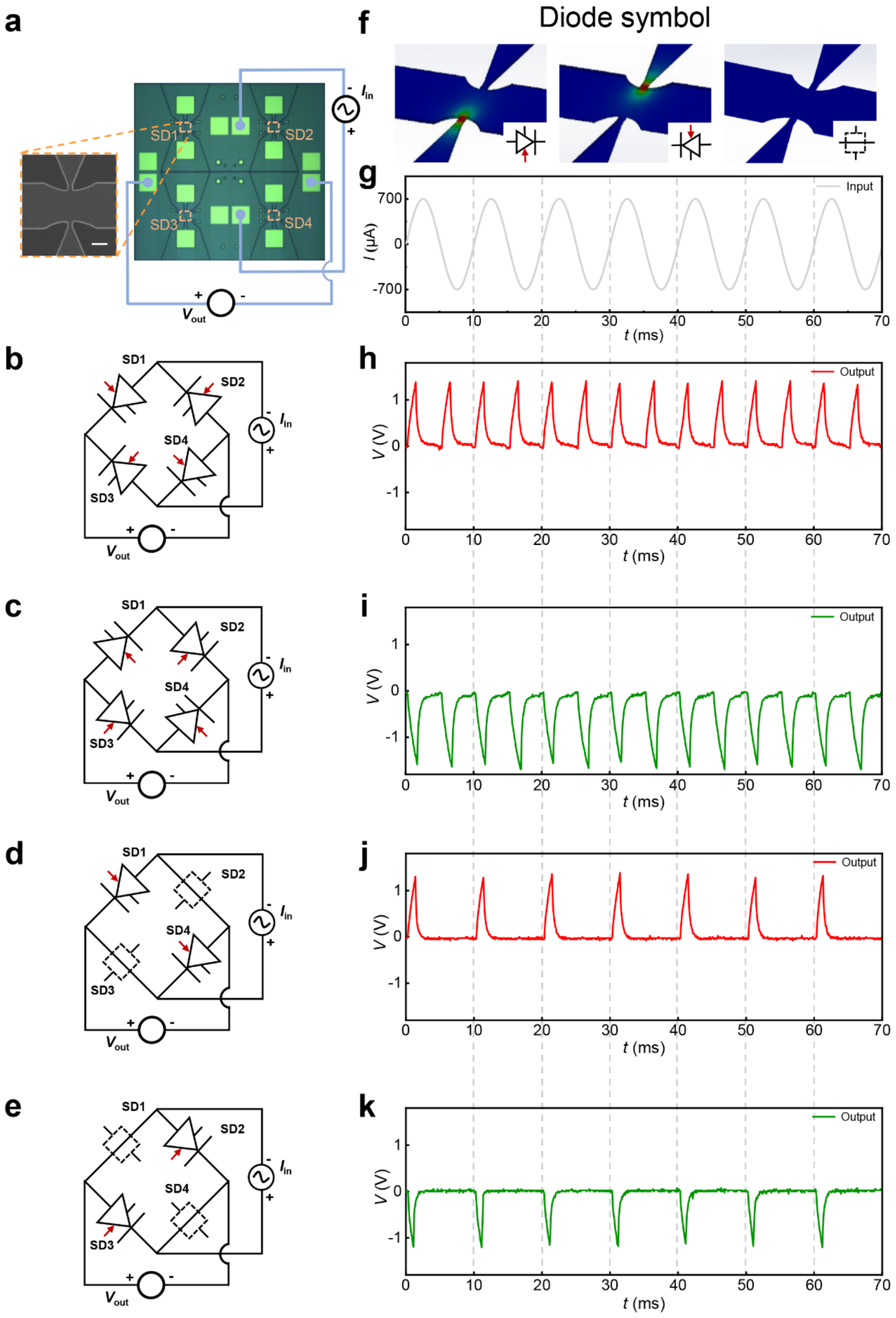

**Figure 4.** Electrically programmable superconducting electric circuit. (a) Optical microscope image of a bridge circuit composed of four identical electrothermal-switch diodes (orange dashed boxes). An SEM image of the upper-left diode is shown on the left. Scale bar, 150 nm. (b)-(e) Circuit diagrams showing different diode configurations of the superconducting bridge in (a). (f) Circuit symbols representing the electrothermal-switch diodes with reversed polarities (left and middle) and as a deactivated superconducting wire (right); the red arrows indicate the gate currents. (g) Sinusoidal input signal with a frequency of 100 Hz. (h)-(k) Measured output rectifications corresponding to the circuit diagrams in (b)-(e): (h) and (i), full-wave rectifications with opposite polarities; (j) and (k), half-wave rectifications with opposite polarities.

The circuit rectification is demonstrated at 100 Hz, a frequency constrained by the low-pass filters in our cryogenic measurement setup—necessary for noise reduction in transport experiments—rather than by any fundamental limitation of the device physics. Based on the material properties and the nTron-like operating mechanism, we estimate the rectification bandwidth could extend to the gigahertz range (see Supporting Information for details). Additionally, to facilitate assessment of the device potential in programmable superconducting circuits, Supporting Information provides an analysis of the steady-state gate power, switching energy, and operational dissipation.

We have demonstrated a nanoscale electrothermal-switch superconducting diode in which a gate-controlled localized hotspot enables in situ and reversible breaking of spatial inversion symmetry in a superconducting nanowire. This electrothermal mechanism gives rise to two distinct diode effects—a nonreciprocal superconducting-to-normal transition and a vortex-motion-induced diode effect—both governed by a unified physical origin. By combining electrical transport measurements with time-dependent Ginzburg–Landau simulations, we reveal how localized electrothermal asymmetry simultaneously modulates vortex entry barriers and superconducting depairing, establishing a microscopic framework for electrothermally driven nonreciprocal superconductivity. The ability to electrically switch the diode polarity and functionality without modifying the magnetic field highlights electrothermal control as a versatile and scalable strategy for engineering superconducting nonreciprocity. More broadly, this electrothermal-switch platform may enable new opportunities for energy-efficient superconducting logic and electrically reconfigurable components in quantum information systems.

## Supporting Information

Supporting Information is available from Online or from the author.

Videos S1 and S2 showing results of G-L simulations for unconventional superconducting vortex diode effects with inversion antisymmetry and antisymmetry breaking (MP4, MP4).

Methods for sample fabrication, experiments and G-L simulations, Supplemental Figures S1-S19，Supplemental Tables S1-S4.

## Notes

The authors declare no competing financial interest.

## Acknowledgements

This work was supported by the Quantum Science and Technology-National Science and Technology Major Project (Grant No. 2024ZD0301300), the National Key R&D (Grant Nos. [2021YFA0718802, 2024YFA1408900]), the National Natural Science Foundation of China (Grant Nos. [62274086, 62288101, 62571230, 92565201, 12504137]), the China Postdoctoral Science Foundation (Grant Nos. [2024M751370, 2025T180930]), the Jiangsu Funding Program for Excellent Postdoctoral Talent (Grant No. 2023ZB534), Natural Science Foundation of Jiangsu Province (Grant No. BK20241223).

# Supplementary information for

## Nanoscale electrothermal-switch superconducting diode for electrically programmable superconducting circuits

Tianyu Li[1,2,#], Jiong Li[3,#], Chong Li[1,4,#], Peiyuan Huang[1,5], Nuo-Zhou Yang[1,2], Wuyue Xu[1,2], Wen-Cheng Yue[1,5], Yang-Yang Lyu[1,*], Yihuang Xiong[1], Xuecou Tu[1,2], Tao Tao[5], Xiaoqing Jia[1,2], Qing-Hu Chen[3,*], Huabing Wang[1,2,*], Peiheng Wu[1,2], and Yong-Lei Wang[1,2,5,*]

[1] *Research Institute of Superconductor Electronics, Nanjing University, Nanjing 210023, China*

[2] *Purple Mountain Laboratories, Nanjing, 211111, China*

[3] *Zhejiang Key Laboratory of Micro-Nano Quantum Chips and Quantum Control, School of Physics, Zhejiang University, Hangzhou 310027, China*

[4] *School of Microelectronics, Nanjing University of Science and Technology, Nanjing, Jiangsu 210094, China*

[5] *State Key Laboratory of Spintronics Devices and Technologies, Nanjing University, Suzhou, China*

# These authors contributed equally

* Correspondence to: yylyu@nju.edu.cn; qhchen@zju.edu.cn; hbwang@nju.edu.cn; yongleiwang@nju.edu.cn

**This PDF file includes:**

Methods, Figures S1 to S19, Tables S1-S4, Videos S1 to S2

# Methods

## Sample fabrication

The electrothermal-switch superconducting diodes were fabricated on $Si/SiO_2$ substrates. First, a 10 nm-thick NbN superconducting film was deposited on a 7.5 × 7.5 $mm^2$ $Si/SiO_2$ chip by magnetron sputtering in a mixed argon and nitrogen atmosphere. The NbN layer was then patterned using ultraviolet lithography followed by reactive ion etching. Subsequently, Ti/Au (10/80 nm) electrodes were defined by ultraviolet lithography and deposited via magnetron sputtering. Finally, the core nanowire structure was patterned by electron beam lithography using polymethyl methacrylate (PMMA A2) resist, followed by $CF_4/SF_6$-based RIE to define the fine device features. The superconducting bridge circuits were fabricated using the same processes as the single-device of the electrothermal-switch diodes.

## Characterization and measurements

The morphology of the superconducting diode was imaged using field-emission scanning electron microscopy (FE-SEM, JEOL-7500F) operated at 1 kV. The electrical transport measurements were performed using the standard four-probe method. A Keithley 6221 current source provided excitation signals, including sinusoidal AC and DC currents. The corresponding DC voltages were recorded using a Keithley 2182A nanovoltmeter, while AC voltages were measured with a Tektronix TBS2000 oscilloscope. The sample was mounted in a superconducting-magnet cryostat (CMI), with the magnetic field oriented perpendicular to the sample plane.

## Time-dependent Ginzburg–Landau (G–L) simulations

The TDGL framework was employed to model the dynamic evolution of the superconducting order parameter and vortex configurations[1,2]. In this model, a superconductor is described by a complex order parameter $\psi$, where $|\psi|^2$ represents the local Cooper-pair density. In dimensionless units and under the zero-electric-potential gauge, the TDGL equations are expressed as:

$$\partial_t \psi = [(\boldsymbol{\nabla} - i\boldsymbol{A})^2 + 1 - T - |\psi|^2]\psi + \chi(\boldsymbol{r}, t) \quad (1)$$

$$\kappa^2 \boldsymbol{\nabla} \times \boldsymbol{\nabla} \times \boldsymbol{A} = \mathrm{Im}\psi^*(\boldsymbol{\nabla} - i\boldsymbol{A})\psi - \sigma \partial_t \boldsymbol{A} \quad (2)$$

Here, $\kappa$ is the Ginzburg-Landau (G-L) parameter, $\sigma$ is the normal-state conductivity, $T$ is the temperature, $\chi(\boldsymbol{r}, t)$ is a random perturbation describing quantum fluctuations, and $\boldsymbol{A}$ is the vector potential, which determines the magnetic field via $\boldsymbol{\nabla} \times \boldsymbol{A} = \boldsymbol{H}$. The time scale is set by the GL relaxation time, $\tau = 4\pi\sigma\kappa^2\xi(0)^2/c^2$, where $c$ is the speed of light. All lengths are normalized to the zero-temperature coherence length $\xi(0)$; temperature to the critical temperature $T_c$; magnetic field to the upper critical field $H_{c2} = \hbar c/[2e\xi(0)^2]$, and the current density in units of $j_0 = \hbar\sigma/[2e\xi(0)\tau]$.

Because the influence of local temperature (Joule heating) on the electrothermal-switch superconducting diode is a key focus of this work, Joule heating was incorporated via the heat-transfer equation：

$$\nu\partial_t T = \zeta\nabla^2 T + \sigma^{-1} j_n^2 + \eta(T_0 - T) \tag{3}$$

where $T_0$ is the bath temperature, $\nu$= 0.03 is the heat capacity, and $\zeta = 0.06$ is the thermal conductivity of the film. The parameter $\eta = 2 \times 10^{-4}$ quantifies the heat-exchange efficiency between the film and its substrate. These phenomenological coefficients correspond to intermediate heat-removal conditions and have been validated as appropriate for typical experimental systems[3,4].

For an ultrathin superconducting film, the simulations were performed in two dimensions at a finite temperature of $T_0 = 0.30T_c$. To match experimental parameters, a strip geometry of length $L_x = 400\xi(0)$ and width $L_y = 100\xi(0)$ was considered, without loss of generality. Representative values of $\kappa = 40$ and $\xi(0) = 5nm$ were used for NbN[5]. The external magnetic field $\boldsymbol{H}_{\mathbf{ext}}$ was applied along the $z$-axis, perpendicular to the film, with a magnitude of $0.008H_{c2}$. A local heat source of width $8\xi(0)$ and temperature $T_s = 3.0T_c$ was placed at the center of the lower boundary. Periodic boundary conditions were applied along the $x$-direction, while Neumann boundary conditions were used along $y$-direction. The external current was introduced via a current-induced magnetic field $H_I$, such that $H|_{y=0} = H_{\text{ext}} - H_I$ and $H|_{y=L_y} = H_{\text{ext}} + H_I$. Equations (1-3) were solved using a semi-implicit algorithm described in Ref.2. The program was implemented in C++ with Compute Unified Device Architecture (CUDA) parallelization on an NVIDIA GeForce RTX 3090 GPU, enabling efficient computation of the evolution across all discretized grid points.

## 1. Further explanation of a spatial thermal gradient that breaks inversion symmetry

We conducted experiments using different gating currents which modifies the Joule heating of the hotspots (Figure S7). As the gate current is increased (enhance Joule heating), both $I_{c+}$ and $I_{c-}$ are progressively suppressed (Fig. S7a) with a monotonic decrease in diode efficiency (Fig. S7b).

We have further performed simulations with a fixed external field $H = 0.008H_{c2}$, varying the hotspot temperature at the center of upper boundary.

The applied gate current locally drives the sample into the normal state. In this regime, the local temperature typically exceeds $T_c$, and the peak temperature of the thermal hotspot can be estimated as $T_s - T_0 \sim I_{G2}^2$, with $T_0$ the bath temperature and $T_s$ the source temperature. The thermal diffusion is governed by the equation

$$\nu \partial_t T = \zeta \nabla^2 T + \sigma^{-1} j_n^2 + \eta (T_0 - T),$$

where $\nu = 0.03$ is the heat capacity of the sample, and $\zeta = 0.06$ represents its thermal conductivity. The parameter $\eta = 2 \times 10^{-4}$ quantifies the efficiency of heat exchange between the sample and its holder. These phenomenological coefficients correspond to moderate heat dissipation conditions. The steady-state spatial temperature profile approximately follows:

$$T(r) - T_0 \sim (T_s - T_0) e^{-\sqrt{\eta/\zeta} r}$$

where $r$ is the distance from the hotspot.

As shown in Fig S7c, both $|J_{c\pm}|$ decrease monotonically with increasing temperature of the heat source. The sample width is $100\xi(0)$, the factor $e^{-\sqrt{\eta/\zeta} r}$ at the lower boundary is on the order $10^{-3}$. Therefore, the potential barrier at the lower boundary is not directly destroyed by the heat source. When a reverse current is applied, the critical current is mainly determined by the current density at the lower boundary, where the transport current adds to the Meissner current. In general, the threshold current density required for a boundary to sustain superconductivity approximately follows the scaling law $\left(1 - \frac{T}{T_c}\right)^{\frac{3}{2}}$.

By contrast, when a forward current is applied, the critical current is mainly governed by vortex motion. Compared with temperature, vortex dynamics are more strongly influenced by the external magnetic field. This can also be inferred from the behavior of depinning current: when the temperature is sufficiently high, the current density required to initiate vortex motion becomes nearly temperature independent. Under these conditions, $J_{c+}$ can be regarded as being mainly related to the width of the barrier-destroyed region (i.e., the normal-state region induced by the heat source), which approximately scales as $\sim\ln(T_s - T_0)$. Consequently, as the temperature increases, $|J_{c-}|$ decays with temperature more rapidly than $|J_{c+}|$. As a result, the diode efficiency decreases with increasing heat-source temperature, until the entire sample becomes overheated by the heat source, at which point the diode effect disappears. These results consist well to experimental data in Figs. S7a and S7b.

## 2. The influence of magnetic fields on the device

Figure S8a presents the *I–V* characteristics measured at zero magnetic field and zero gate current at various temperatures below the superconducting transition temperature (2–7 K). In this case, the positive and negative critical currents are nearly identical, indicating negligible nonreciprocity in the absence of both magnetic field and gate-induced electrothermal asymmetry.

We further measured the *I–V* characteristics at zero magnetic field with the gate currents ($I_{G2}$ = +10 μA, $I_{G2}$ = −10 μA, and $I_{G1}$ = −10 μA) (Figure S8b). The results show that even when an electrothermal hotspot is generated by the gate current, the positive and negative critical currents remain nearly symmetric in zero field. These results suggest that the time-reversal symmetry breaking induced by a magnetic field is necessary to the nonreciprocity, consistent to the general understanding of superconducting diode effect.

An applied magnetic field generates vortices and produces opposing Meissner screening currents at the two edges of the nanowire. Consequently, the rectification efficiency becomes field-dependent (Fig. 3), consistent with the mechanism proposed in Ref. 10 (see Comment #2 of Reviewer #1). In particular, the vortex-mediated SDE (V-SDE) relies on the presence of magnetic-field-induced vortices and therefore disappears at zero magnetic field. Under finite magnetic fields, vortices are present, and the gate-induced hotspot creates an asymmetric vortex entry barrier, leading to the observed nonreciprocal transport. For a given gate current and applied magnetic field, the rectification efficiency initially increases with magnetic field strength, reaches a maximum, and then decreases as the field increases further (Figure 3). The detailed mechanism of the field dependence is described in Ref. 10.

## 3. The states of each diode in the full-wave rectifier

We use the bridge circuit in Figure S11a as an example and provide a detailed breakdown of the operational status of each diode during different temporal intervals, as illustrated in Figure S11 and summarized in Table 1.

i) Fully superconducting interval (Green). When the input sinusoidal signal is in the region marked green (Figure S11b), all four diodes remain in the superconducting state. Consequently, no voltage develops across the bridge output

ii) Mixed-state interval (Grey). As the signal enters the grey region, the diodes exhibit mixed states: SD2 and SD3 stay superconducting, while SD1 and SD4 transition to the normal state. This configuration, corresponding to the grey region in Figure S11c, enables rectification of the input waveform.

iii) Fully normal interval (Pink). All four diodes were drived into the normal state. In this interval, represented by the pink region in Figure S11c, the rectification ceases.

## 4. Dependence on gate current amplitude

To elucidate the origin of the nonreciprocal transport, we examined the effect of gate current magnitude on the diode behavior. As shown in Figure S13, the degree of nonreciprocity gradually diminishes with increasing gate current amplitude. This trend can be attributed to the enhanced electrothermal homogenization induced by larger gate currents, which suppresses the spatial temperature gradient responsible for breaking inversion symmetry. These observations provide additional evidence supporting the thermal-gradient-mediated mechanism proposed in the main text.

## 5. Temperature dependence of nonreciprocal transport

We next investigated the temperature dependence of the SDE. The magnetic-field-dependent critical currents were measured over a temperature range of 2 K to 7.5 K (Figure S14). The results show that the device exhibits clear nonreciprocity of the critical current ($\pm I_c$) below its superconducting transition temperature $T_c$, with the nonreciprocal magnitude decreasing monotonically as temperature increases. The strongest nonreciprocity occurs at $T$ = 2 K, indicating that the diode behavior is intrinsically linked to superconducting coherence and electrothermal feedback. The corresponding diode efficiency was also calculated as a function of temperature (Figures S15 and S16). The efficiency decreases with increasing temperature, reaching a peak value at $T$ = 2 K. These results confirm the robustness and universality of the superconducting diode effect across the entire superconducting regime.

## 6. Maximum diode efficiency at different temperatures

To provide a quantitative overview, the maximum diode efficiencies obtained at different temperatures are summarized in Figure S17. The results reveal a gradual reduction in the maximum efficiency as temperature rises. This correlated evolution of diode efficiency further supports the thermally driven origin of the nonreciprocal behavior.

## 7. Correction for asymmetry in normal-state resistance

A minor discrepancy was observed between the normal-state resistances measured directly under positive and negative current biases (Figure S18a). This asymmetry arises from the presence of the gate current, which slightly offsets the effective bias current. If the current in the positive direction is defined as $+I_1$ and the current in the opposite direction is $-I_1$, the applied gate current is $I_0$, and the sample current measured by the ammeter is $I_1'$, then there is a relationship:

$$+I_1' = I_1 + I_0 \tag{1}$$

$$-I_1' = -I_1 + I_0 \tag{2}$$

At this time, $\triangle I_1' = 2I_0$, which results in the artificial inconsistency of normal resistance.

To address this, we use the following method to correct this effect: normalize the normal state resistances together, find the effective value of the biasing gating current influence, and then subtract this value (half of it) from the data of both positive and negative currents. The formula is as follows：

$$R_+ = R_- = \frac{|V_+|}{I_+ + \frac{|I'|}{2}} = \frac{|V_-|}{I_- + \frac{|I'|}{2}} \tag{3}$$

## 8. The bandwidth limitation and switching speed of the electrothermal-switch superconducting diode

From the practical measurement perspective, the present rectification demonstrations at 100 Hz are limited by our cryogenic measurement setup rather than by the intrinsic device physics. In our current low-temperature system, each measurement line uses enamel-coated copper wires and includes RC low-pass filters with a cutoff frequency of approximately ~1 kHz (In Supplemental Fig. S9, we expanded our diode operation frequency up to 1313 Hz). This configuration is optimized for low-noise DC and low-frequency transport measurements, which is necessary for nanowire measurements. While it is well suited for characterizing low-frequency nonreciprocal transport, it does not preserve high-frequency signal integrity. Therefore, the experimentally demonstrated 100 Hz operation reflects instrumental constraints. To address the reviewer's question regarding the intrinsic bandwidth, we discuss the expected limits based on reported references and thermal mechanisms.

(1) Comparison with references

From previously reported superconducting diode devices based on microbridges of different materials and geometries, the maximum demonstrated operating frequencies are summarized in Table S3.

These reports indicate that the achievable operating frequency of superconducting microbridge diodes typically lies in the kHz~MHz range, depending strongly on device material and geometry. Compared with previously reported superconducting diode microbridges, our device features a smaller bridge width, a thin film thickness, and intrinsically fast electron-phonon relaxation in NbN. These factors are expected to enhance thermal dissipation efficiency and reduce the hotspot relaxation time. In particular, compared with the NbN microbridge in [*Nat. Electron.* 8, 417 (2025)], our device has a smaller bridge width and therefore a reduced thermal volume. Based on scaling considerations, we conservatively estimate that the intrinsic operating frequency of our device could reasonably reach 100 MHz level, and potentially higher with optimized thermal engineering.

(2) Thermal dynamics

Beyond empirical comparison, the intrinsic frequency limitation in electrothermal-switch-type devices is primarily governed by thermal formation and relaxation during diode switching. The relevant microscopic processes include electron-phonon coupling time ($\tau_{e\text{-}ph}$) and phonon escape time to the substrate ($\tau_{esc}$). For thin-film NbN, typical reported values are ~10 ps for $\tau_{e\text{-}ph}$ and ~100 ps for $\tau_{esc}$ [*Phys. Rev. B* 102, 054501 (2020)]. The thermal formation is governed by electron-phonon coupling, leading to an estimated hotspot formation time on the order of ~10 ps. In contrast, the thermal relaxation time is mainly governed by phonon escape time to the substrate, yielding on the order of ~100 ps in NbN nanobridges. This corresponds to an intrinsic thermal bandwidth of approximately $f_{max} = 1 / (2\pi\tau_{esc}) \approx 1.6$ GHz. Notably, similar NbN nanowires in superconducting nanowire single-photon detectors (SNSPDs) routinely operate at detection rates in the hundreds of MHz to GHz regime [*APL Photonics* 10, 106113 (2025)], further supporting that NbN nanowires can sustain ultrafast electrothermal dynamics. Considering practical non-idealities and readout limitations, a conservative operating frequency around ~1 GHz is reasonable.

(3) Superconducting applications

Regarding the intended applications, the estimated intrinsic bandwidth is fully compatible with many superconducting devices and subsystems, including detectors such as SQUIDs and SNSPDs, which typically operate below GHz regime. Also, superconducting diodes have recently been proposed as key nonreciprocal elements for coherent control and signal routing in circuit quantum electrodynamics architectures, enabling functionalities such as direction-dependent qubit coupling and nonreciprocal quantum gates [arXiv:2511.20758 (2025)]. In addition, high-efficiency superconducting diodes have been suggested as building blocks for superconducting digital logic circuits, where logic gates such as NOT, AND, OR, NAND, and NOR can be realized by exploiting tunable diode polarity [arXiv:2410.23352 (2024)]. Therefore, the GHz-level intrinsic bandwidth estimated for our device is sufficient for emerging superconducting electronics and quantum processing applications, with electrically tunable nonreciprocity enabling programmable superconducting circuit elements.

## 9. The energy and dissipation characteristics of the device

(1) Steady-state gate power.

In our device, the steady-state gate power required to maintain the electrothermal hotspot is approximately 42 nW, estimated from the applied gate current and voltage across the gate lead. Most prior reports on electrically tunable superconducting diodes [*Nat. Nanotechnol.* 16, 760 (2021); *Nat. Commun.* 13, 3658 (2022); *Nat. Commun.* 14, 3078 (2023), etc.] do not disclose comparable power metrics, precluding direct quantitative comparisons. Nevertheless, our electrothermal-switch device simultaneously achieves superior rectification efficiencies (up to 60% in the V-SDE regime) and an ultra-compact nanoscale footprint, which is highly advantageous for circuit-level integration.

In addition, we have compared it with other reported electrothermal-switch superconducting devices based on n-Tron, and the order of magnitude is comparable, which typically operate in the nanowatt regime. For clarity, we have compiled representative values from the literature and summarized them in Table 2, showing that the gate power in our device falls within the typical operating range of previously reported n-Tron-type structures. Of course, its energy consumption is related to the specific structural dimensions. The energy consumption we are presenting here is not the optimal value for this method. In the future, the energy consumption can be further reduced by optimizing the materials and structural dimensions.

(2) Energy considerations for switching.

In Fig. S9, the diode switching under a square-wave current. We estimated the power dissipation in the normal-state operation regime and compared it with previously reported magnetically and electrically controlled superconducting diodes. The results are summarized in Table 3. Our device achieves a diode efficiency of ~60%, which is among the highest values for electrically controlled platforms and comparable to the best-performing entries (~70%). Critically, it operates at significantly higher critical currents (±70 μA to +270 μA) than most electrically gated devices (typically nA – few μA range), providing substantially greater current-carrying capacity suitable for practical superconducting circuit integration. This larger operating current naturally leads to a higher normal-state power dissipation of ~135 μW—several orders of magnitude above the pW–nW values of low-current electrically controlled diodes, but still well below the mW level of some magnetically controlled systems and fully consistent with the nanowatt-to-microwatt regime expected for n-Tron-based architectures.

The elevated dissipation is therefore not a fundamental limitation but a direct consequence of the device's design for higher current handling and efficiency. As shown in the Table S3, further optimization remains feasible through material selection and geometric scaling (e.g., reducing nanowire width or hotspot size), which would proportionally lower both normal-state resistance and required bias currents while preserving the electrothermal-switch mechanism and high rectification performance.

(3) Added dissipation in the channel.

The added dissipation introduced during diode operation mainly originates from the localized electrothermal hotspot generated by the gate current, which is responsible for breaking inversion symmetry and enabling nonreciprocal transport. Under steady-state conditions, this dissipation is primarily determined by the gate-induced heating power (~42 nW). The main nanowire channel itself remains superconducting over most of the operating regime, except during the vortex-motion or superconducting-to-normal transition regimes described in the manuscript.

Importantly, the presence of the hotspot lowers the critical current required for the full superconducting-to-normal transition across the channel (see Figure S19). This reduction in switching threshold decreases the overall energy dissipation associated with the transition to the normal state.

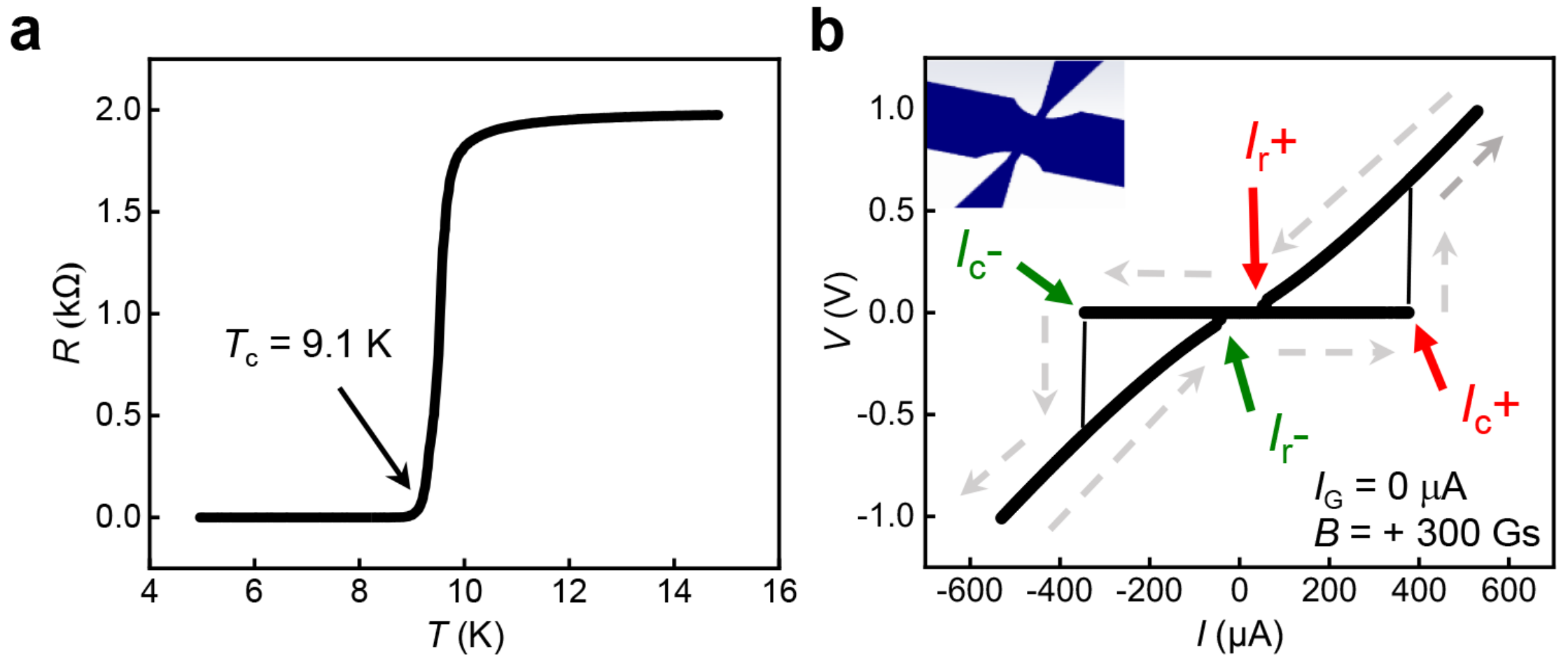

**Figure S1.** Temperature-dependent resistance and symmetric *I*-*V* characteristics of the superconducting nanowire. (a) Temperature-dependent resistance of the superconducting nanowire, showing a critical temperature $T_c$= 9.1 K. (b) *I*-*V* characteristics measured at zero gate current under a magnetic field of 300 Gs. The zero-gate-current curve exhibits symmetric behavior with negligible nonreciprocity. Arrows in the figure indicate the sweep directions of the applied current.

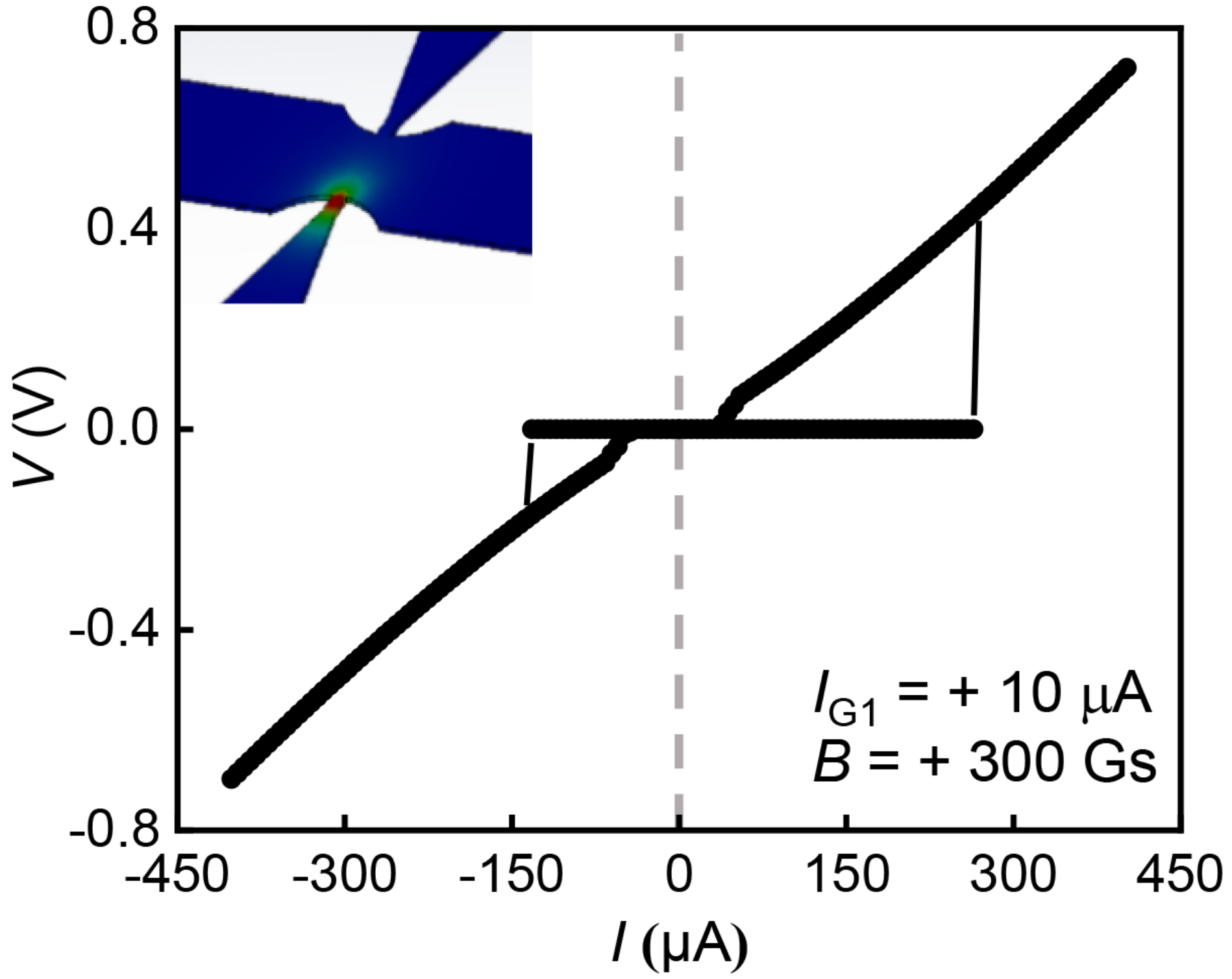

**Figure S2.** *I-V* characteristics of the electrothermal–switch superconducting diode device. *I-V* characteristics measured with a gate current $I_{G1}$= 10 μA applied to one gate lead at 2 K and an external magnetic field of 300 Gs.

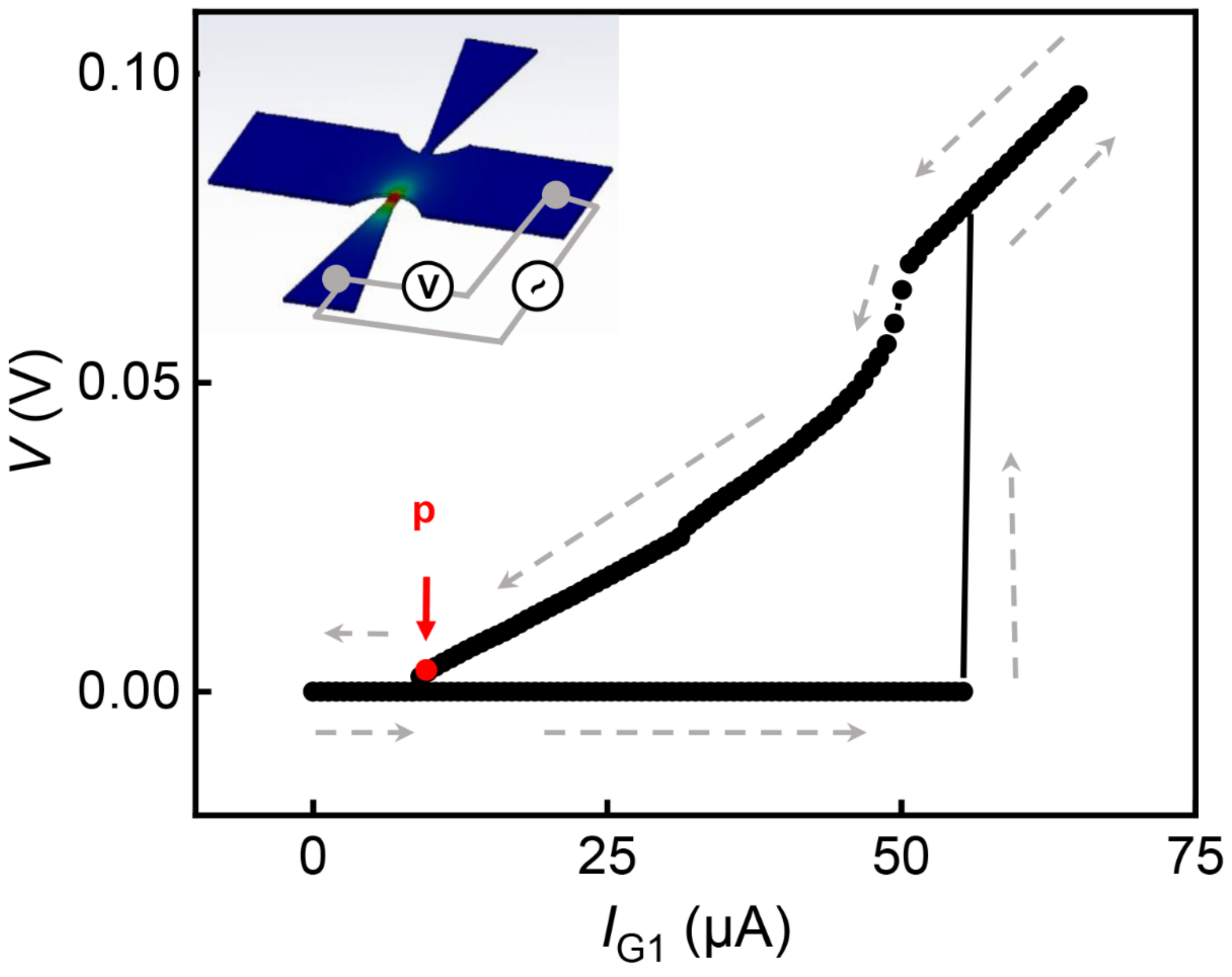

**Figure S3.** *I*-*V* characteristics of the gate lead. *I*-*V* characteristics of the gate lead (The illustration in the upper corner is a schematic of the circuit used for measuring), from which the retrapping current $I_r = 9.12$ μA is determined. The device operates slightly above $I_r$ (red point) to generate a stable nanoscale resistive hotspot without quenching the main superconducting channel.

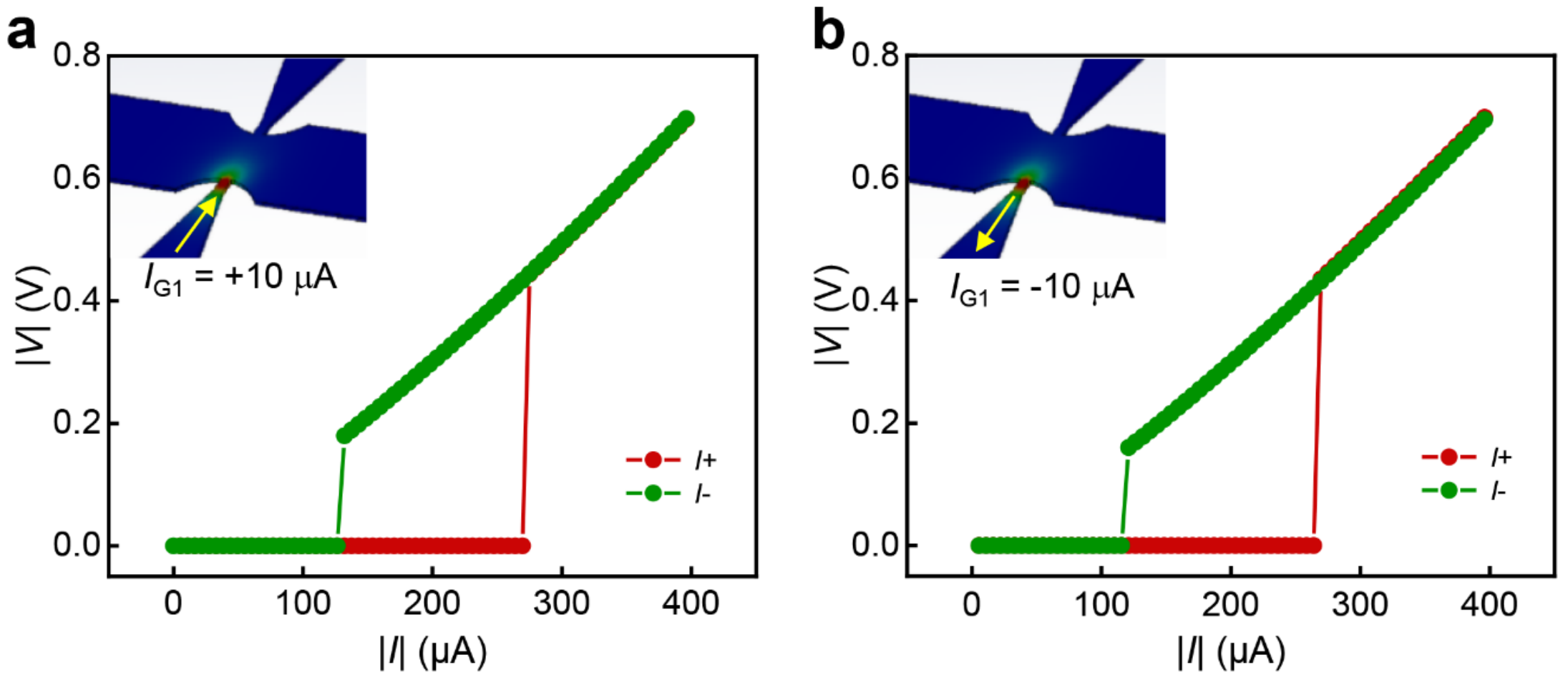

**Figure S4.** Polarity-independent nonreciprocity confirming the electrothermal-switch mechanism. (a) and (b) *I*-*V* characteristics measured with opposite gate currents applied to the same gate lead: $I_{G1}$= +10 μA (a) and $I_{G1}$= -10 μA (b) under a magnetic field of 300 Gs at 2 K. The two results exhibit nearly identical nonreciprocal behavior, indicating that reversing the gate-current direction does not affect the diode polarity or efficiency, which is consistent with the electrothermal-switch mechanism. The insets schematically illustrate that the hotspot position remains unchanged under opposite gate-current directions.

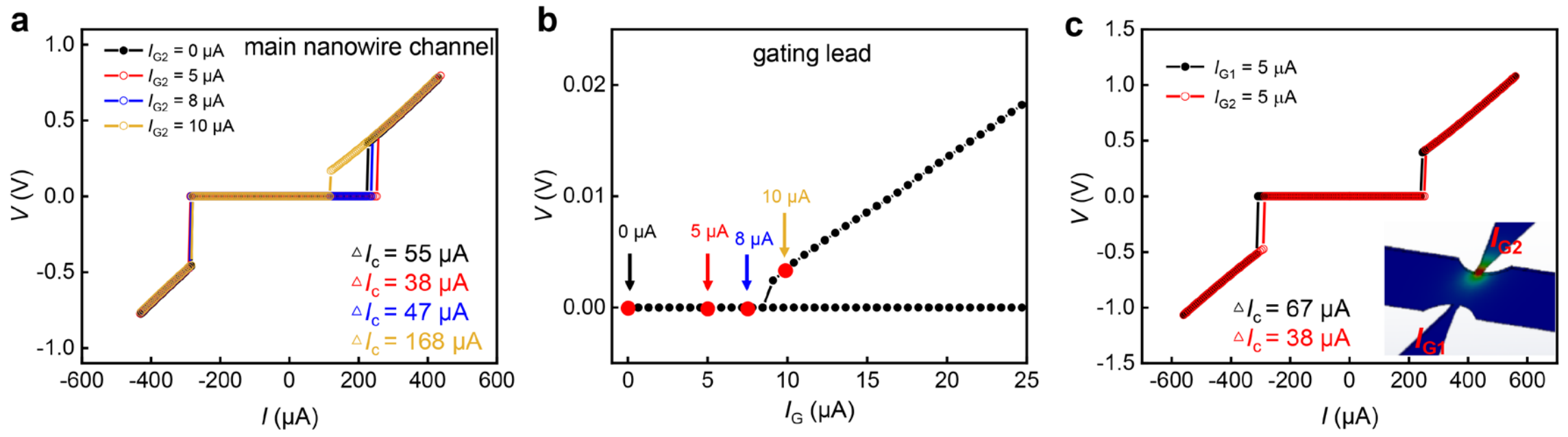

**Figure S5.** Hotspot-enabled superconducting diode effect. (a) *I*-*V* characteristics of the nanobridge measured under the representative gate currents, $I_G$ = 0, 5, 8 μA, and 10 μA and under the magnetic field of 400 Gs and temperature of $T$ = 3 K. (b) Corresponding *I*-*V* curve of the gate electrode. The four data points highlighted in (b) correspond to the gate current values at which *I*-*V* curves in (a) were measured. (c) The *I*-*V* curves obtained under $I_{G1}$ = 5 μA and $I_{G2}$ = 5 μA, respectively ($B$ = 400 Gs, $T$ = 3 K).

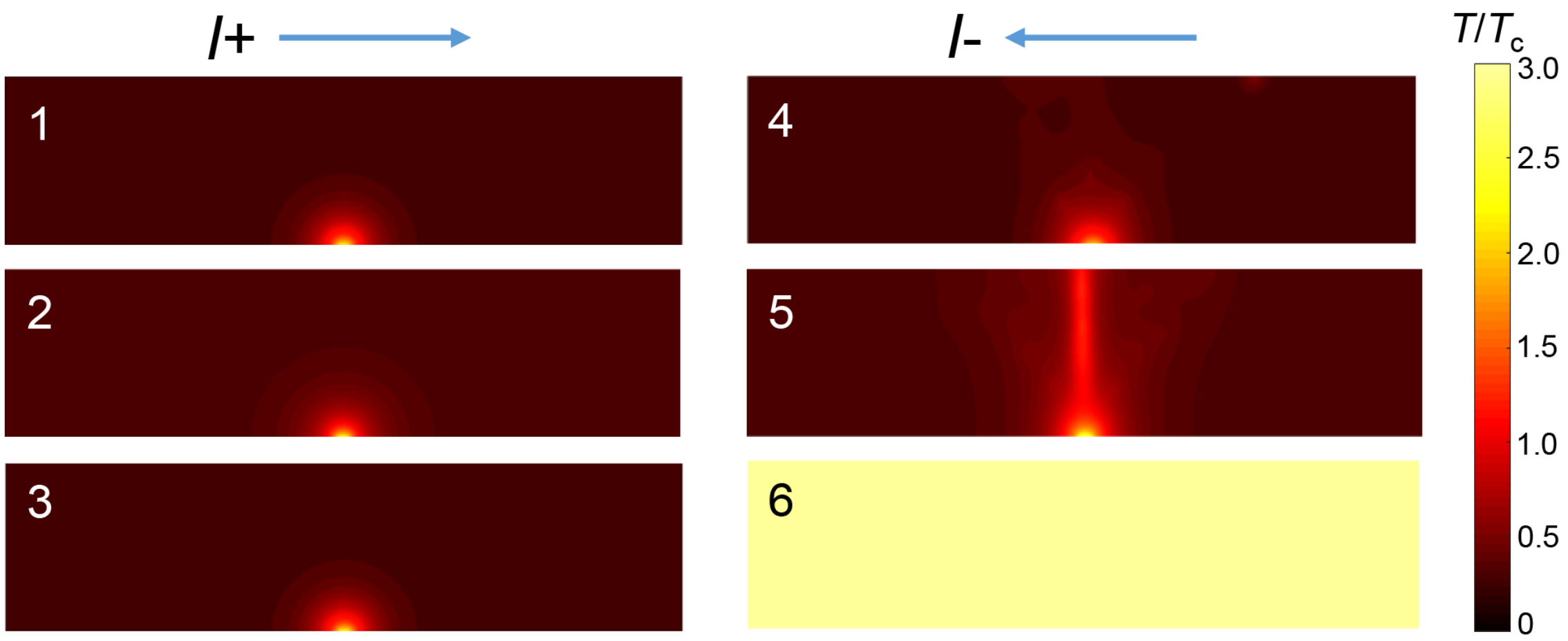

**Figure S6.** Ginzburg–Landau simulations of the thermal-switched superconducting diode. Snapshots from Video S2, obtained from TDGL simulations, illustrating the spatial distribution of the temperature under positive (left) and negative (right) currents. The snapshots correspond to the *I*-*V* points 1-6 indicated in Figure 2(a).

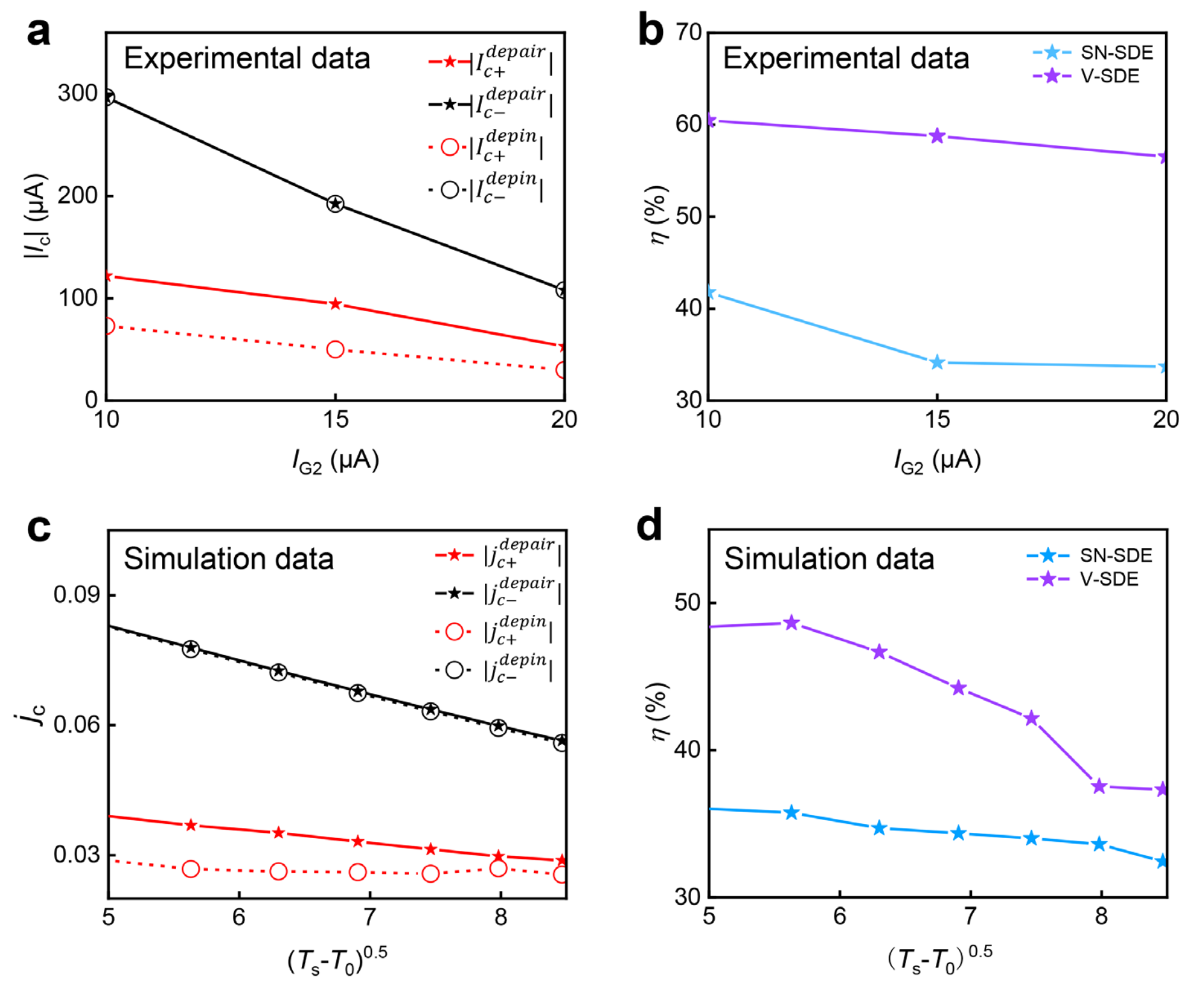

**Figure S7.** Electrothermal control of nonreciprocal transport. At $T = 2$ K and $B = 320$Gs, the extracted positive and negative critical currents (a) and diode efficiency (b) from the *I*-*V* characteristic curves under different sizes of gating current. Under external field $H = 0.008H_{c2}$, the simulated positive and negative critical currents (c) and diode efficiency (d) under varying source temperature at the center of upper boundary.

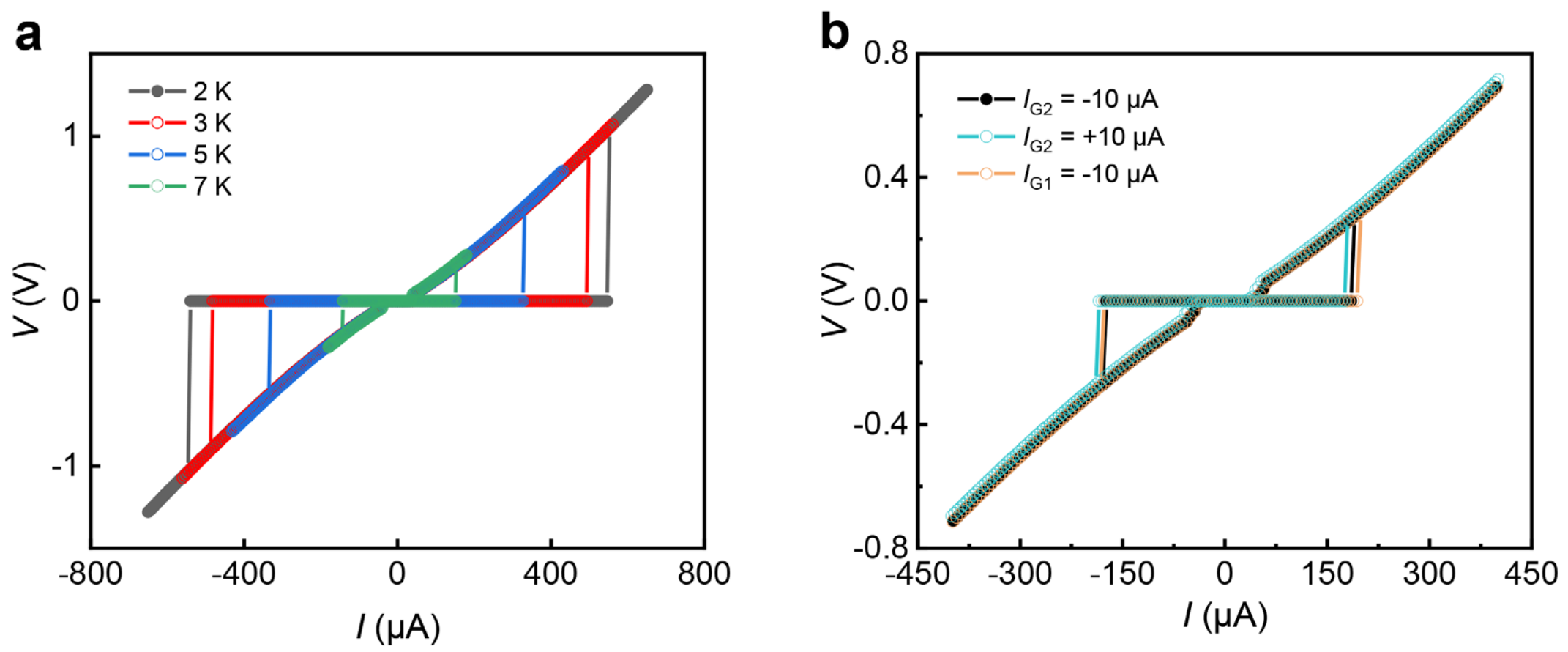

**Figure S8.** Absence of diode effect under zero magnetic field. (a) *I*–*V* characteristics with different temperature ($B$ = 0, $I_G$ = 0). (b) *I*–*V* characteristics with finite gate currents ($B$ = 0).

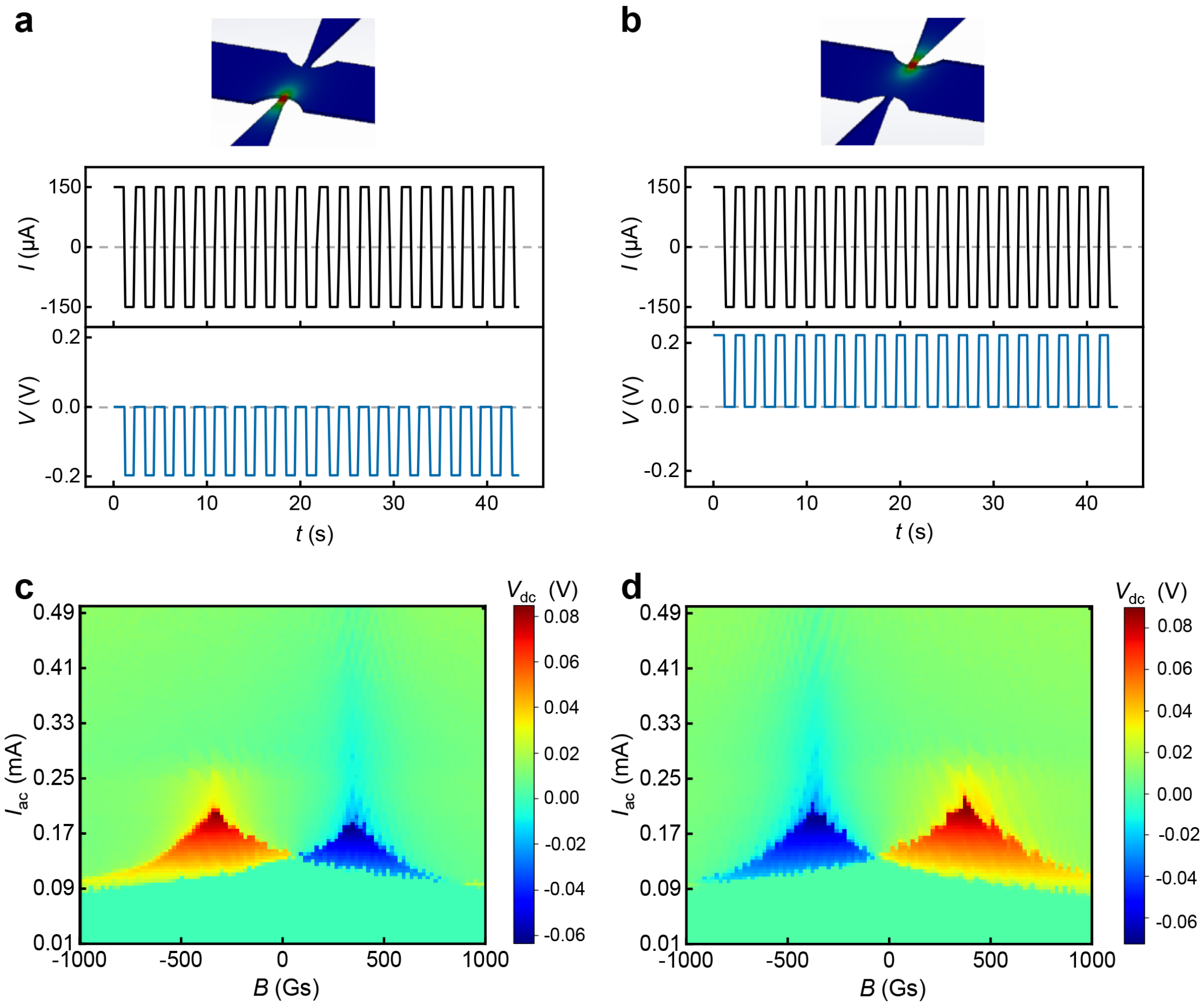

**Figure S9.** Rectification effect of the electrothermal-switch superconducting diode. (a) and (b) Half-wave rectification measured under a square-wave current drive, with gate currents applied to the bottom ($I_{G1}$= 15 μA) (a) and upper ($I_{G2}$= 15 μA) (b) gate leads. Measurements were performed under a magnetic field of 320 Gs at 2 K. (c) and (d), Color maps showing the dependence of the DC rectification voltage on the amplitude of a sinusoidal alternating current (1313 Hz) at various magnetic fields, measured under $I_{G1}$ = 15 μA (c) and $I_{G2}$ = 15 μA (d) at 2 K.

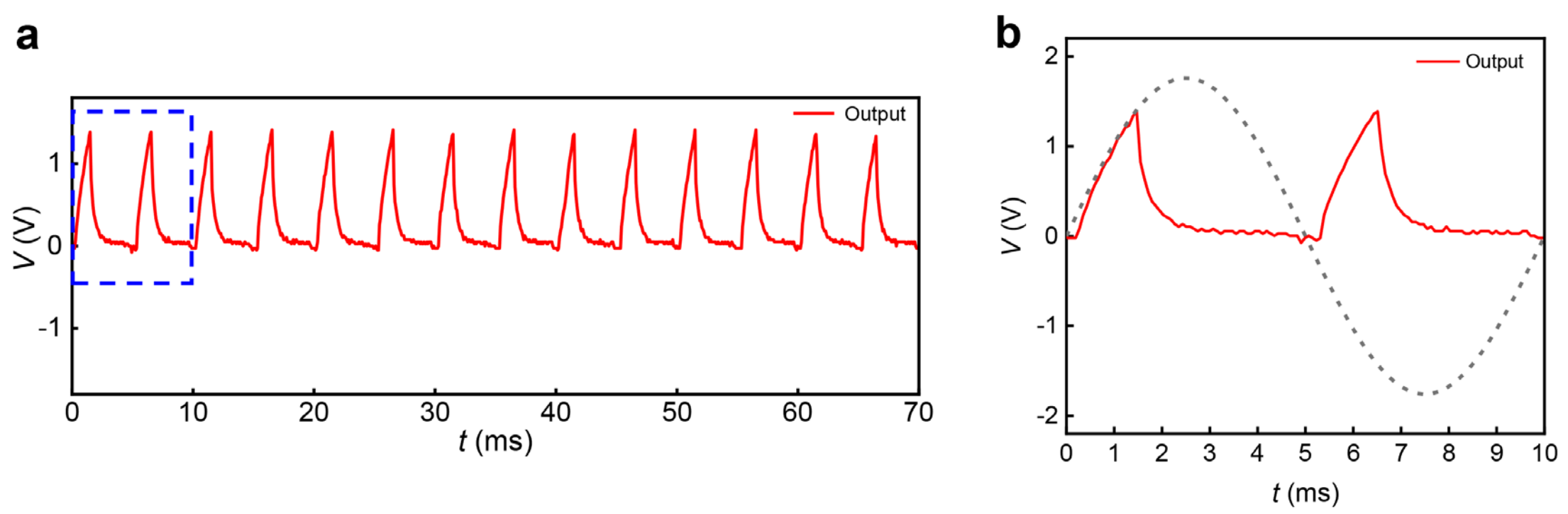

**Figure S10.** Full-wave rectification output. (a) The output rectification of the rectifier under a sinusoidal input signal. (b) Enlarged view of the region selected by the dashed blue box in (a), with the dashed curve showing the time-domain sinusoidal excitation.

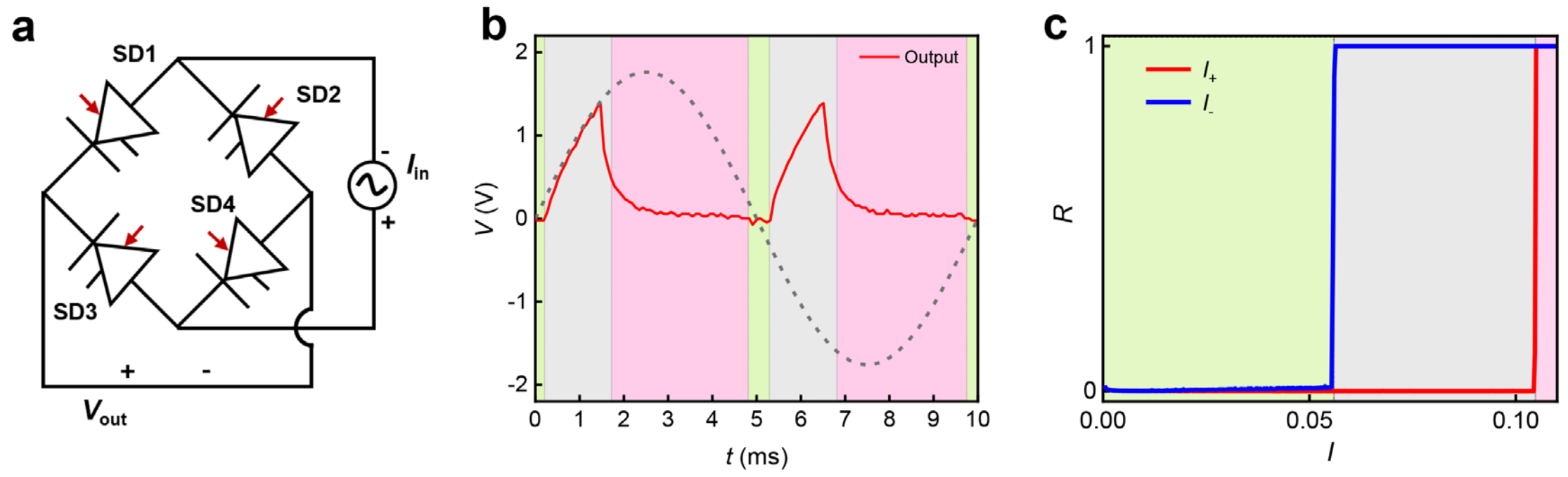

**Figure S11.** Operational states of diodes during rectification. (a) Circuit diagram of the full-wave rectifier. (b) Measured output rectified signal under a 100 Hz sinusoidal input (single period shown). (c) Resistance–current (*R*–*I*) characteristics illustrating the diode states corresponding to the colored regions in (b).

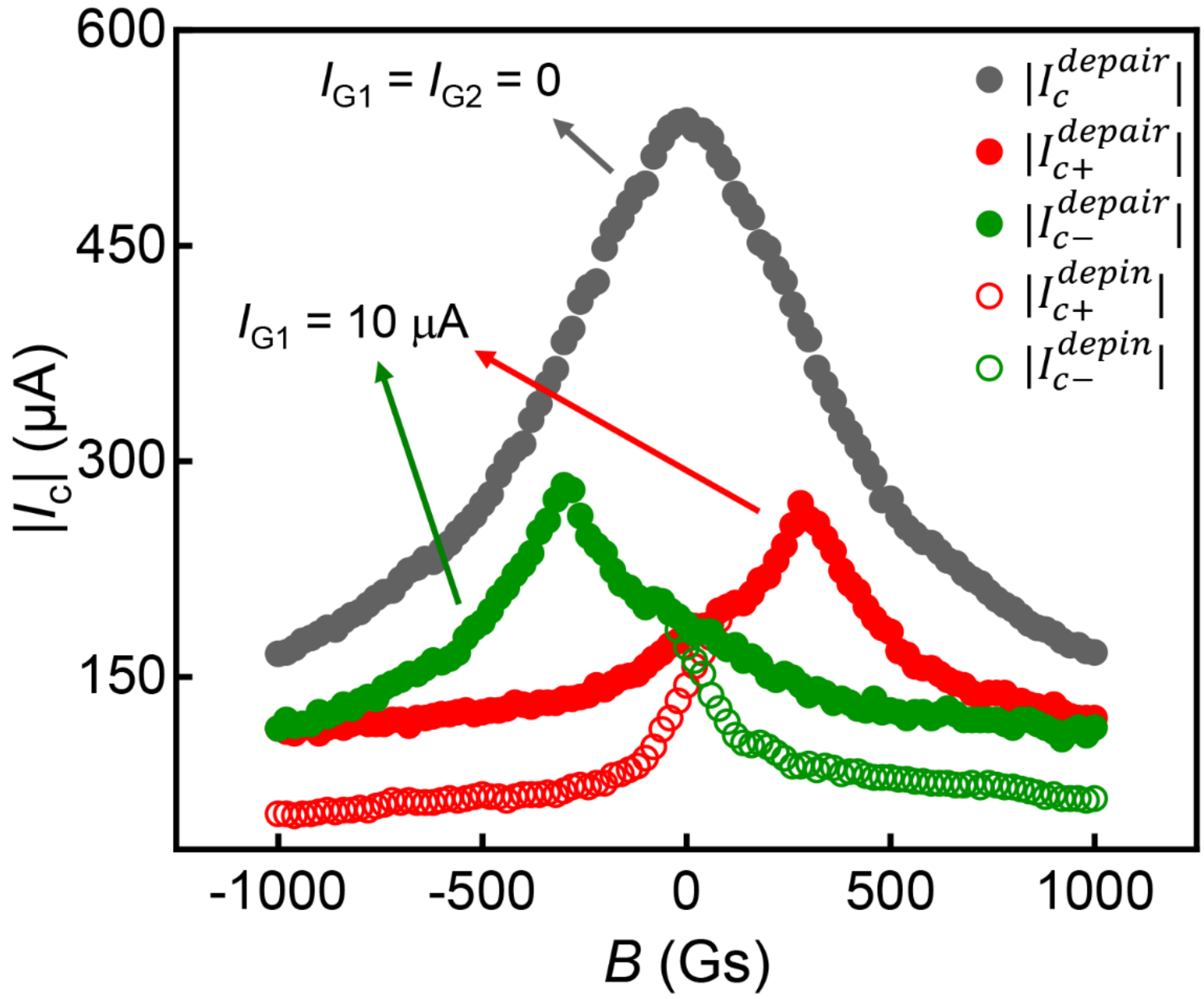

**Figure S12.** Comparison of the critical currents with and without a gate current. The gate current generates a localized hotspot in the nanowire channel, suppressing the critical transition currents across the entire magnetic field range. As a result, the critical currents of the zero-gate-current configuration are consistently higher than those of the current-gated device, regardless of the current directions.

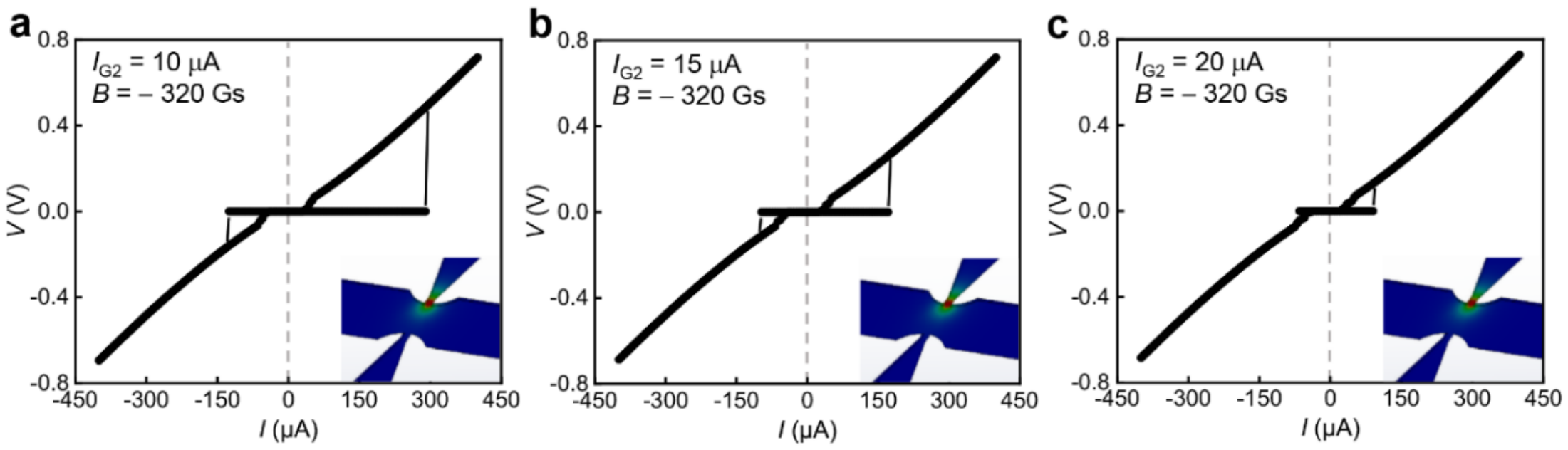

**Figure S13.** Gate-current dependence of the *I*-*V* characteristics. *I*-*V* curves of the device were measured at $T = 2$ K and $B = -320$ Gs, under different gate currents: (a) $I_{G2} = 10$ μA; (b) $I_{G2} = 15$ μA; (c) $I_{G2} = 20$ μA.

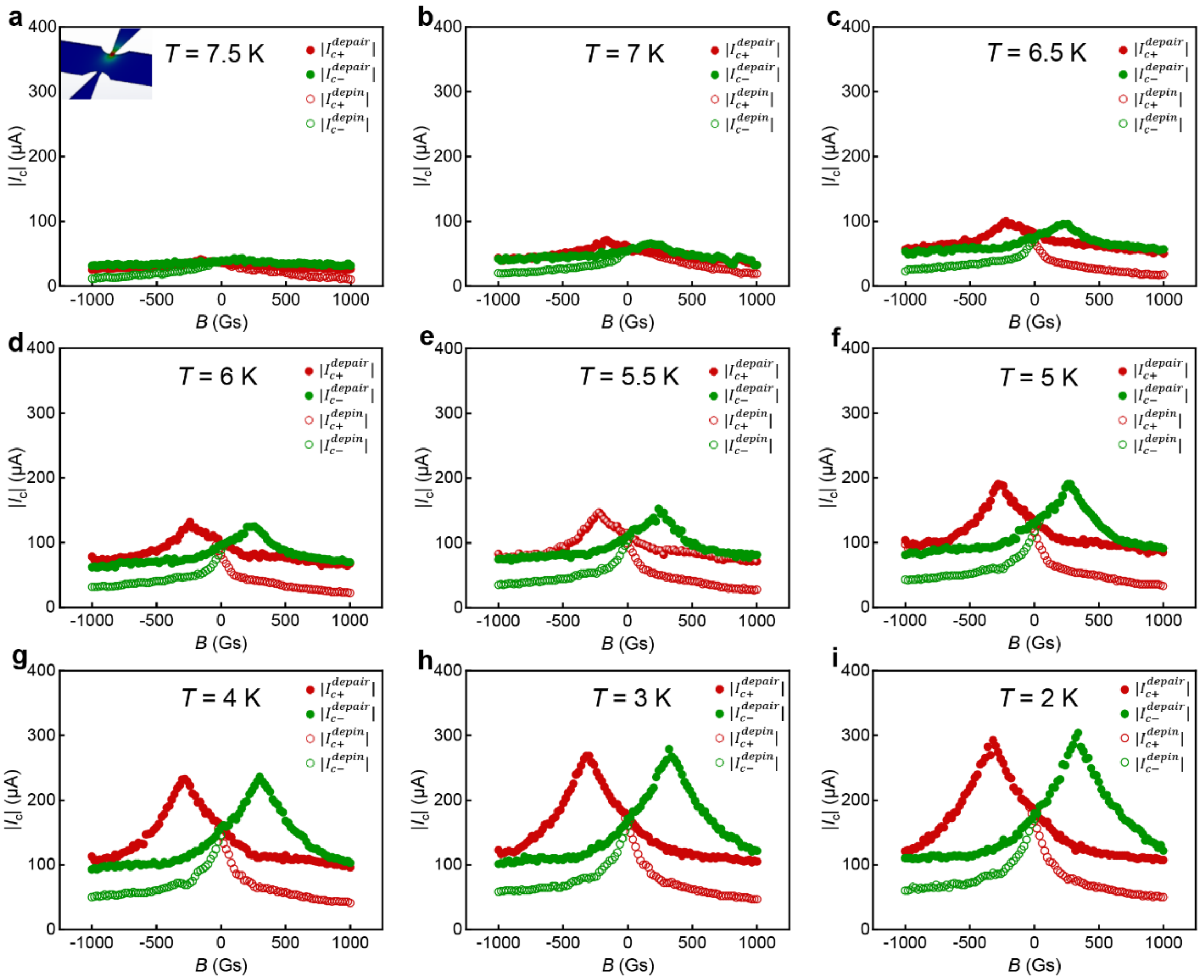

**Figure S14.** Magnetic field dependences of nonreciprocal critical current. Magnetic-field dependences of positive and negative superconducting critical currents for V-SDE and SN-SDE were measured at different temperatures: (a) 7.5 K; (b) 7 K; (c) 6.5 K; (d) 6 K; (e) 5.5 K; (f) 5 K; (g) 4 K; (h) 3 K; (i) 2 K. All measurements were performed under a fixed gate current of $I_{G2}$= +10 μA.

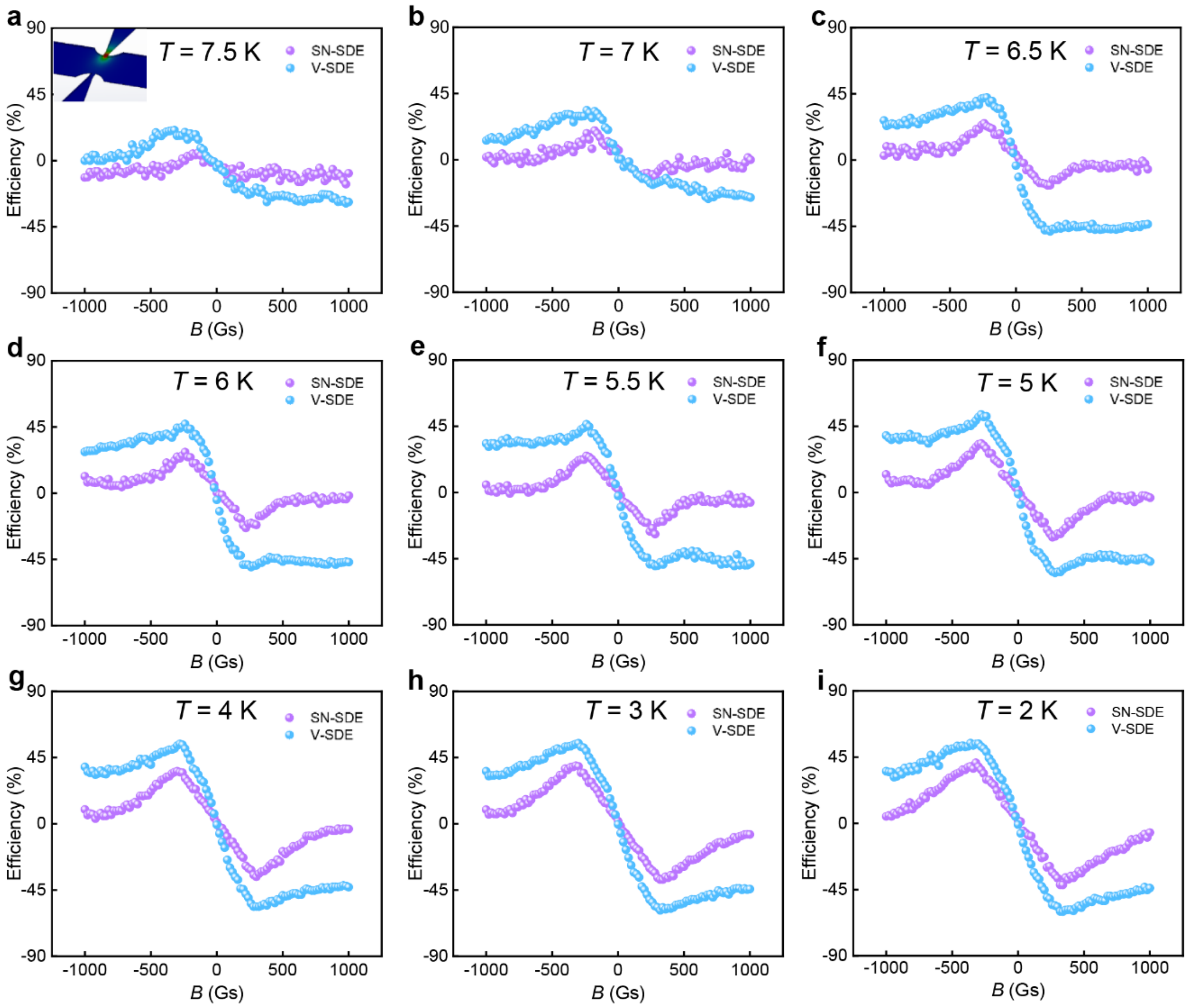

**Figure S15.** Temperature dependence of diode efficiency. Diode efficiencies were calculated at different temperatures, corresponding to the same device shown in Figure S14.

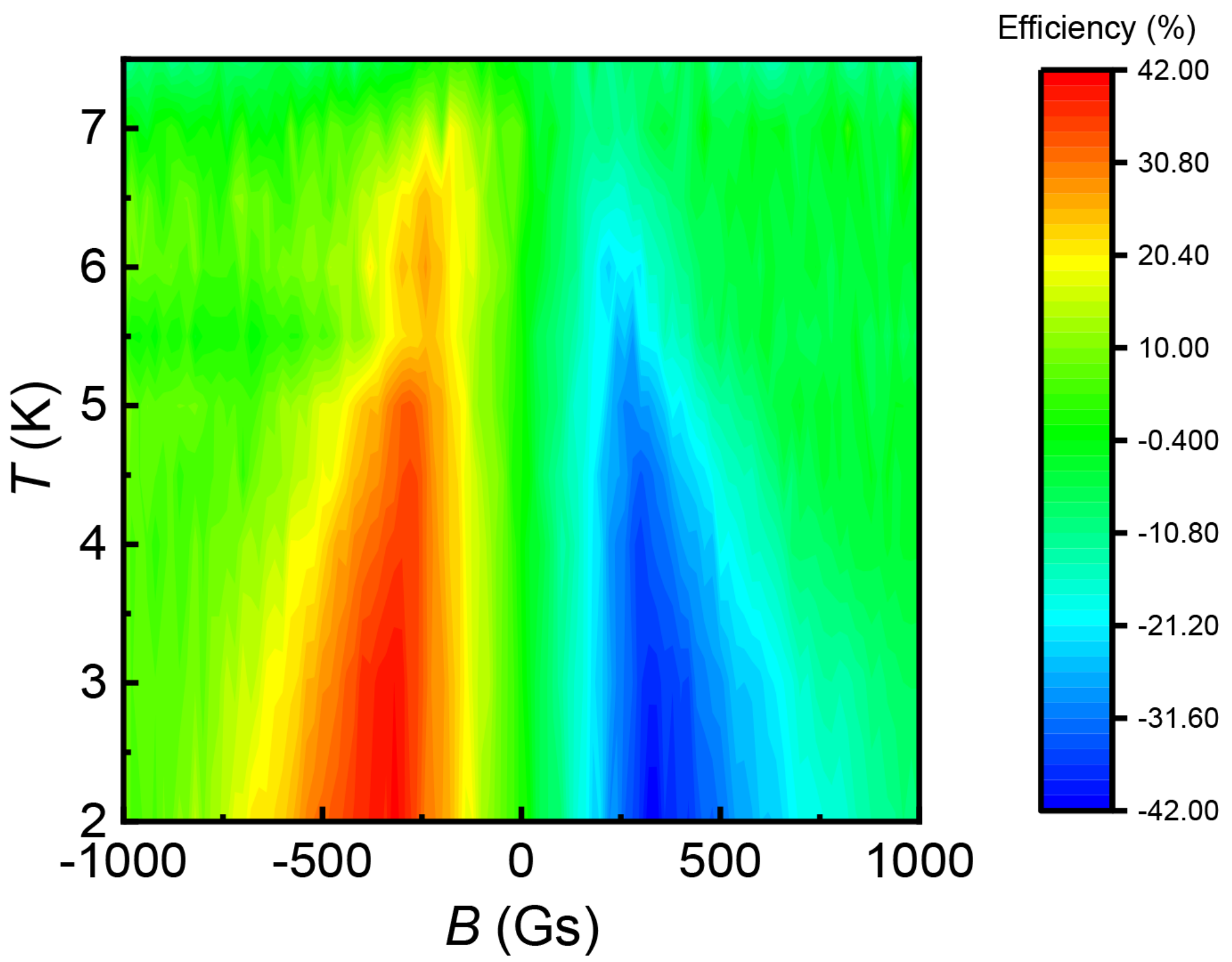

**Figure S16.** Colormap of diode efficiency. Diode efficiencies were obtained at different temperatures and magnetic fields, corresponding to the same device shown in Figure S14.

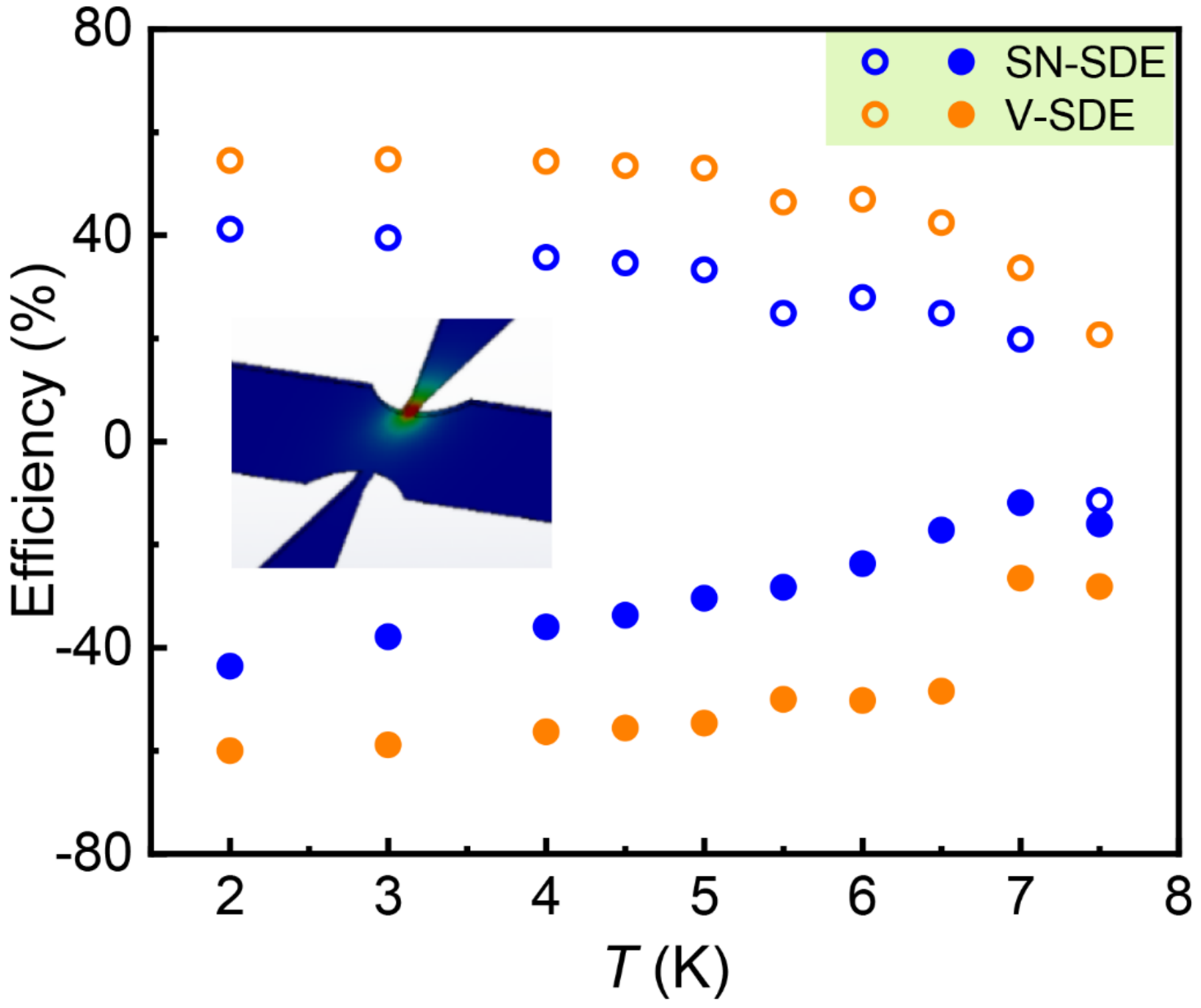

**Figure S17.** Maximum diode efficiency as a function of temperature. Summary of the maximum diode efficiencies obtained by the device at different temperatures.

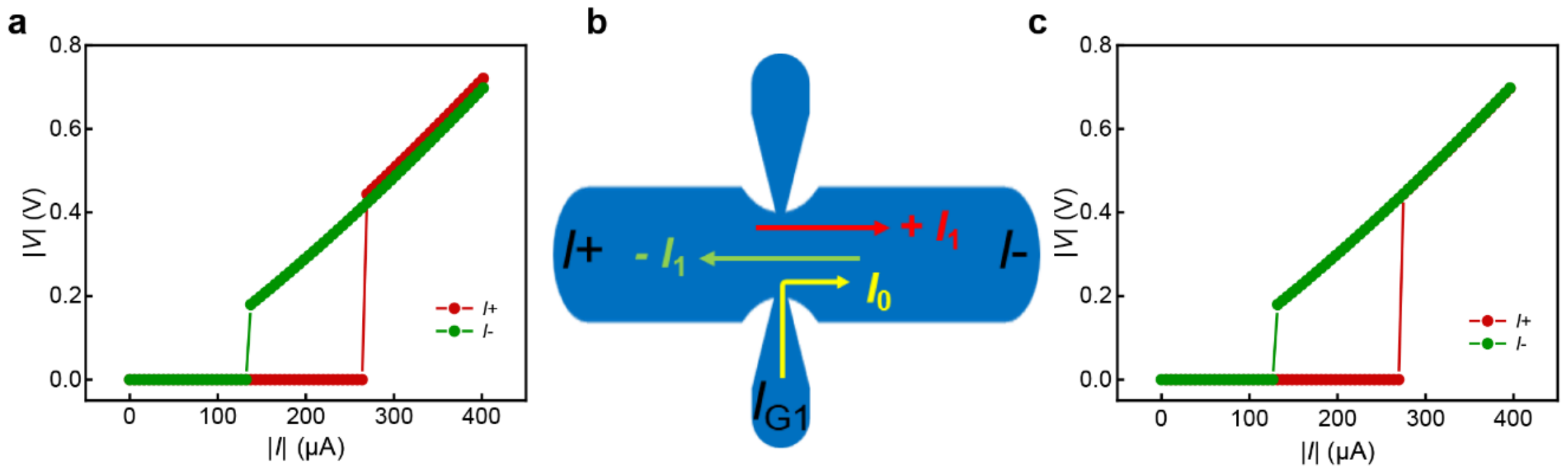

**Figure S18.** Correction of the asymmetry in normal-state resistance. (a) Example showing the slight discrepancy in the normal-state resistances of the *I*-*V* curves for positive and negative bias currents. (b) Schematic illustration explaining the origin of the resistance asymmetry caused by the gate-current offset. (c) Corrected *I*-*V* characteristics after normalization and subtraction of the gate-induced bias component, demonstrating the restored symmetry in the normal-state resistance.

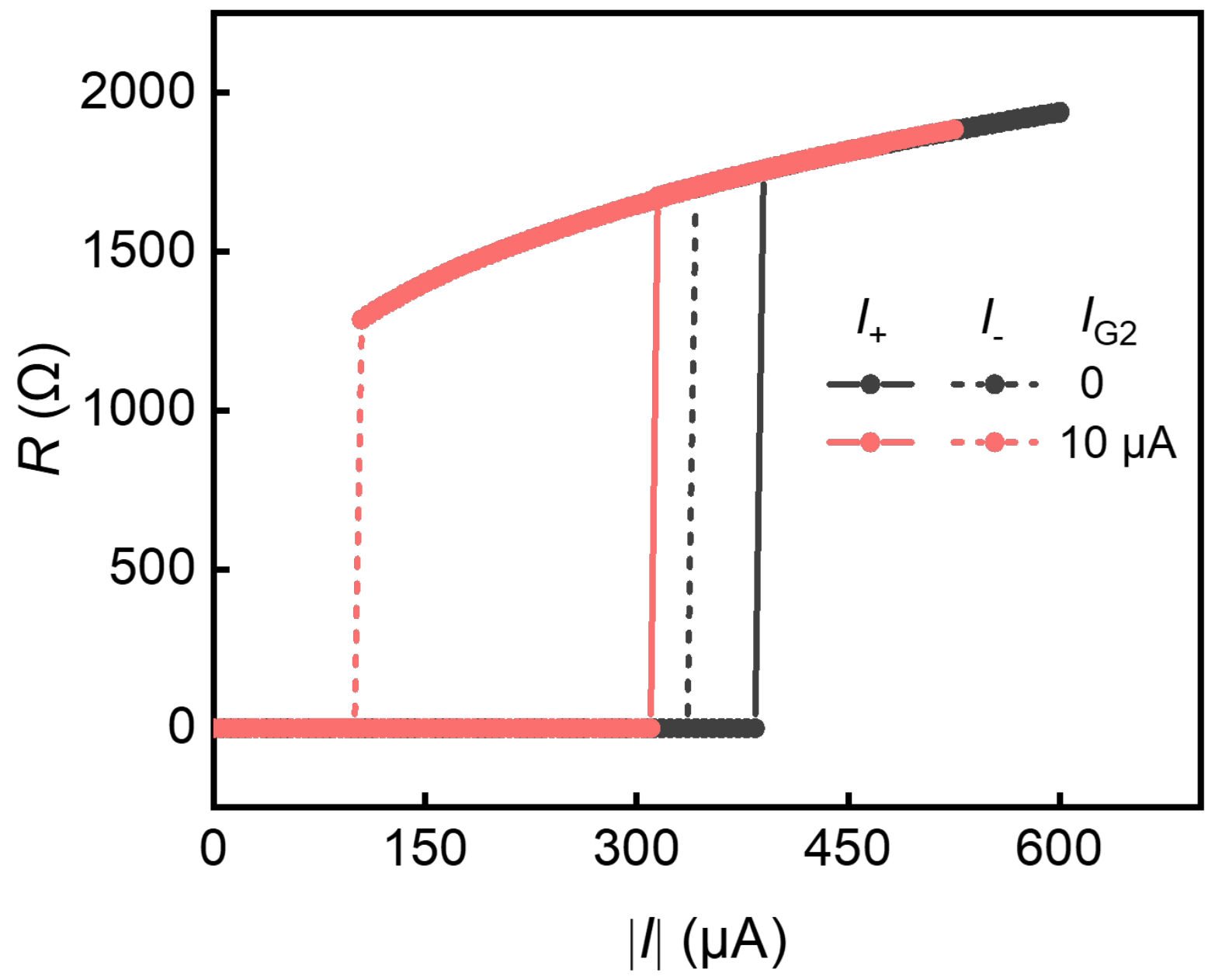

**Figure S19.** The *I*-*V* characteristic curves measured with and without gating under $B$ = -320 Gs.

| Input / SD state / SD | Green region | Grey region | Pink region |
|---|---|---|---|
| **SD1** | Zero-resistance | Normal-resistance | Normal-resistance |
| **SD2** | Zero-resistance | Zero-resistance | Normal-resistance |
| **SD3** | Zero-resistance | Zero-resistance | Normal-resistance |
| **SD4** | Zero-resistance | Normal-resistance | Normal-resistance |

**Table S1.** The state of the four diodes in the circuit shown in Figure S11.

| Reference | Type | Energy |
|---|---|---|
| *Superconductor Science and Technology* 30, 044002 (2017) | Electrothermal-switch n-Tron | ~ 8 nW |
| *Scientific Reports* 9, 16345 (2019) | Electrothermal-switch n-Tron | ~ 19.7 nW |
| *Nano Lett.* 20, 3553 (2020) | Electrothermal-switch n-Tron | ~ 1 nW |
| *Appl. Phys. Lett.* 122, 092601 (2023) | Electrothermal-switch n-Tron | ~ 182 nW |
| Our work | Electrothermal-swtich SC diode | ~ 42 nW |

**Table S2.** Comparison of steady-state gate power based on n-Tron structure devices.

| Reference | Normal resistance | $\pm I_c$ range | Diode efficiency | Magnetically/ Electrically control | Energy |
|---|---|---|---|---|---|
| *Nat. Commun.* 13, 4266 (2022) | 1 kΩ | -9 $\mu$A – +5 $\mu$A | ~30% | Magnetically control | ~ 70 nW |
| *Adv. Funct. Mater.* 34, 2311229 (2023) | ~1.5 Ω | -15 $\mu$A - +50 $\mu$A | ~50.4% | Magnetically control | ~ 14 nW |
| *Adv. Quantum Technol.* 7, 2300378 (2024) | ~12 kΩ | -25 $\mu$A - +40 $\mu$A | ~25% | Magnetically control | ~ 12 nW |
| *Phys. Rev. B* 107, 054506 (2023) | ~25 Ω | -10 *m*A - +11 *m*A | ~10% | Magnetically control | ~ 6 mW |
| *Nat. Commun.* 13, 3658 (2022) | 30 Ω | -40 $\mu$A - +200 $\mu$A | ~70% | Electrically control | ~ 2.5 nW |
| *Nat. Nanotechnol.* 16, 760 (2021) | 4 kΩ | -6*n*A - +8 *n*A | ~15% | Electrically control | ~ 0.4 pW |
| *Nat. Commun.* 14, 3078 (2023) | ~100 Ω | -0.6 $\mu$A - +0.8 $\mu$A | ~15% | Electrically control | ~ 50 pW |
| Our work | 2 kΩ | -70 $\mu$A - +270 $\mu$A | ~60% | Electrically control | ~ 135 μW |

**Table S3.** Comparison of power dissipation for tunable superconducting diodes.

| Reference | Material | Width * Thickness | Working Frequency |
|---|---|---|---|
| *Nat. Electron.* 8, 417 (2025) | NbN | 1 μm * 10 nm | ~ 100 MHz |
| *Nat. Electron.* 8, 411 (2025) | V/EuS | 8 μm * 8 nm | ~ 40 kHz |
| *Nat. Commun.* 12, 2703 (2021) | MoGe | 50 μm * 50 nm | ~ 30 kHz |
| *Phys. Rev. B* 107, 054506 (2023) | $Nb_3Sn$ | 5 μm * 100 nm | ~100 kHz |
| Our work | NbN | 0.5 μm * 10 nm | / |

**Table S4.** Comparison of superconducting diode operating frequencies.

**Video S1.** Evolution of the superconducting order-parameter distribution. The left panel shows the simulated *I-V* characteristics, with the y-axis divided into two linear scales to clearly reveal both the low- and high-dissipation regimes. The right panel illustrates the corresponding spatial evolution of the superconducting order parameter under positive (upper) and negative (lower) current directions.

**Video S2.** Evolution of the temperature distribution. The left panel shows the simulated *I-V* characteristics, with the y-axis divided into two linear scales to clearly reveal both the low- and high-dissipation regimes. The right panel displays the corresponding temperature distributions under positive (upper) and negative (lower) current directions.